\documentclass[10pt,journal,compsoc]{IEEEtran}

\ifCLASSOPTIONcompsoc

  \usepackage[nocompress]{cite}
\else
  \usepackage{cite}
\fi

\ifCLASSINFOpdf
  \usepackage[pdftex]{graphicx}
  
\else
  
\fi

\usepackage{amsmath}

\usepackage{url}

\usepackage{amssymb}

\usepackage{algorithm,algpseudocode,float}

\usepackage{lipsum}

\makeatletter
\newenvironment{breakablealgorithm}
  {
   \begin{center}
     \refstepcounter{algorithm}
     \hrule height.8pt depth0pt \kern2pt
     \renewcommand{\caption}[2][\relax]{
       {\raggedright\textbf{\ALG@name~\thealgorithm} ##2\par}%
       \ifx\relax##1\relax 
         \addcontentsline{loa}{algorithm}{\protect\numberline{\thealgorithm}##2}%
       \else 
         \addcontentsline{loa}{algorithm}{\protect\numberline{\thealgorithm}##1}%
       \fi
       \kern2pt\hrule\kern2pt
     }
  }{
     \kern2pt\hrule\relax
   \end{center}
  }
\makeatother

\usepackage{multirow}
\usepackage{multicol}

\usepackage{color}

\begin{document}

\title{\scriptsize{This work has been submitted to the IEEE for possible publication. Copyright may be transferred without notice, after which this version may no longer be accessible.\\}\huge{Design for Assurance: Employing Functional Verification Tools for Thwarting Hardware Trojan Threat in 3PIPs}}

\author{Wei~Hu,~\IEEEmembership{Member,~IEEE},
	    Beibei~Li,~\IEEEmembership{Student Member,~IEEE},
            Lingjuan~Wu*,
            Yiwei~Li,\\
            Xuefei~Li,
            and~Liang~Hong~\IEEEmembership{Member,~IEEE}
\IEEEcompsocitemizethanks{\IEEEcompsocthanksitem W. Hu, L. Hong, B. Li, Y. Li and X. Li are with the School of Cybersecurity, Northwestern Polytechnical University, Xi'an 710072,  China. E-mail: \{weihu, hongliang\}@nwpu.edu.cn; \{beibeili, lywei, hdfflxf1997\}@mail.nwpu.edu.cn.\protect\\
\IEEEcompsocthanksitem L. Wu is with the College of Informatics, Huazhong Agricultural University, Wuhan 430070, China. E-mail: wulj@mail.hzau.edu.cn.}}

\IEEEtitleabstractindextext{%
\begin{abstract}
Third-party intellectual property cores are essential building blocks of modern system-on-chip and integrated circuit designs. However, these design components usually come from vendors of different trust levels and may contain undocumented design functionality. Distinguishing such stealthy lightweight malicious design modification can be a challenging task due to the lack of a golden reference. In this work, we make a step towards design for assurance by developing a method for identifying and preventing hardware Trojans, employing functional verification tools and languages familiar to hardware designers. We dump synthesized design netlist mapped to a field programmable gate array technology library and perform switching as well as coverage analysis at the granularity of look-up-tables (LUTs) in order to identify specious signals and cells. We automatically extract and formally prove properties related to switching and coverage, which allows us to retrieve Trojan trigger condition. We further provide a solution to preventing Trojan from activation by reconfiguring the confirmed malicious LUTs. Experimental results have demonstrated that our method can detect and mitigate \emph{Trust-Hub} as well as recently reported don't care Trojans.
\end{abstract}

}

\maketitle

\IEEEdisplaynontitleabstractindextext

\IEEEpeerreviewmaketitle

\IEEEraisesectionheading{\section{Introduction}
\label{sec:intro}}
\IEEEPARstart{M}{odern} system-on-chip (SoC) and integrated circuit (IC) designs typically involve integration of third-party intellectual property (3PIP) cores to accelerate the design process and shorten time-to-market. These IP components are usually delivered by vendors of different trust levels and may contain undocumented functionality known as hardware Trojan~\cite{karri2010trustworthy,Tehranipoor2010Taxonomy}. When activated, Trojans can create covert channels and attack surfaces to leak secret information~\cite{hu2017sdc}, escalate privilege~\cite{Yang2016A2Trojan}, downgrade performance~\cite{Zhao2018DoS} or even induce physical damage~\cite{Li2016Destructible}. If such malicious design modifications remain undetected during the design phase, they would reside like logic bombs in personal devices~\cite{Swierczynski2017}, critical infrastructures~\cite{Gunti2017SCADA} and even military defense systems~\cite{Adee2008Hunt,skorobogatov2012breakthrough}.

Hardware Trojans are usually carefully crafted lightweight circuitries, which stay dormant most of the time and are triggered only under rare events. The possibility of accidentally covering the trigger condition within the exponential-scale design state space during limited testing approaches \emph{zero}. Without an effective solution, searching for a few additional malicious function units out of millions of design cells is no different from looking for a needle in the vast sea. Although the research community has developed various hardware Trojan detection techniques, such as functional~\cite{Salmani2012Reducing,zhang2015veritrust} and security verification~\cite{hu2016detecting,Guo2019Trojan}, switching probability analysis~\cite{waksman2013fanci}, side channel analysis (SCA)~\cite{Narasimhan2013MultipleParameter,Huang2018Scalable} and reverse engineering~\cite{Bao2016Reverse}, existing solutions fall short in several aspects. Specifically, functional and security verification based techniques heavily rely on the quality of test vectors and assertion properties~\cite{Deutschbein2022Property}. In addition, they may miss Trojans hiding behind don't care conditions~\cite{hu2017sdc,fern2015hardware,hu2019leveraging}, which do not cause violation of specification. Switching probability analysis based methods can see high false positive rates and may fail to detect Trojans with multiple discrete triggers~\cite{hu2017sdc,Zhang2014DeTrust}. SCA based solutions usually require a golden chip and are sensitive to process variations~\cite{Cui2018Variation}. Reverse engineering based approaches require the support of specialized equipment and thus are very expensive.

Recently, there has been numerous interest in computer aided design (CAD) for assurance~\cite{Takarabt2018EDA,Knechtel2020Secure,CADWeb}. It is desirable to develop hardware security verification and quantitative analysis solutions, which leverage standard hardware description languages (HDLs), property specification languages (PSLs) and electronic design automation (EDA) tools, allowing designers without rich security knowledge to automate the secure hardware design process. Such efforts will significantly narrow the semantic gap between functional and security verification methodologies, which will eventually allow security to be integrated as an additional decision variable in design space exploration and evaluated alongside functional correctness and performance. We make an attempt to enhance the functional CAD flow with security extensions, using hardware Trojan detection and prevention in 3PIPs as a demonstration.

In this article, we propose a field programmable gate array (FPGA) oriented design flow for identifying and preventing hardware Trojans, leveraging functional verification tools familiar to hardware designers. We dump synthesized design netlist mapped to an FPGA technology library and perform switching as well as coverage analysis at the granularity of LUTs in order to identify specious functional units. We then automatically extract and formally prove invariant properties related to switching and coverage to recover Trojan trigger conditions. We further provide a method for preventing Trojan from activation by reconfiguring the identified malicious LUTs. The major differences that distinguish this work from existing hardware Trojan detection solutions lie in that \emph{the proposed method targets LUTs, which offer an ideal granularity for coverage analysis, automated property extraction and design reconfiguration.} Specifically, this work makes the following contributions:
    \begin{enumerate}
    \item[--]{Developing a design flow for pinpointing hardware Trojan designs in 3PIPs through LUT level switching and coverage analysis;}
    \item[--]{Proposing a method to automatically extract Trojan related properties and recover Trojan trigger condition through formal proof;}
    \item[--]{Providing an approach for hardware Trojan prevention through LUT reconfiguration;}
    \item[--]{Presenting experimental results to demonstrate the effectiveness of the proposed method in detecting and preventing hardware Trojans.}
    \end{enumerate}
 
The remainder of this article is organized as follows. In Section~\ref{sec:related}, we review related work in design level Trojans as well as existing techniques for detecting and preventing Trojans in 3PIPs. We discuss our threat model in Section~\ref{sec:threat} and cover some preliminaries in Section~\ref{sec:preliminary}. Section~\ref{sec:flow} introduces our method for detecting hardware Trojans through LUT level switching and coverage analysis, retrieving Trojan trigger condition by proving Trojan related properties and preventing Trojan from activation by reconfiguring the identified malicious LUTs. We present experimental results in detecting and mitigating the \emph{Trust-Hub} as well as recently reported don't care Trojans in Section~\ref{sec:results} and conclude the article in Section~\ref{sec:conclusion}.

\section{Related Work}
\label{sec:related}
Hardware Trojans are oftentimes inserted during the design or fabrication stages. This work focuses on the design phase and aims to develop a design flow for thwarting hardware Trojan threat in 3PIPs. In Section~\ref{sec:tj-design}, we briefly review the related research in design level Trojans. We discuss various techniques for detecting and preventing hardware Trojans in Sections~\ref{sec:detecttech} and \ref{sec:preventtech} respectively.

\subsection{Design Level Hardware Trojans}
\label{sec:tj-design}
Tehranipoor \emph{et al.} \cite{Tehranipoor2010Taxonomy} proposed a comprehensive taxonomy for classifying different hardware Trojans, according to the insertion phase, abstraction level, activation mechanism, effect and location.  It allows the hardware design and security communities to characterize different Trojans and develop highly targeted solutions to detecting and preventing Trojan threats.  Salmani \emph{et al.} further supplemented the classification systems of hardware Trojans and released a complete set of benchmarks for evaluating the effectiveness of different hardware Trojan detection methods~\cite{Salmani2013Trust-Hub}. These benchmarks use simple trigger mechanisms such as a timer or a specific input pattern to generate a single explicit trigger signal, which can be easily identified by switching probability analysis methods~\cite{waksman2013fanci}. To evade existing Trojan detection techniques, newer Trojan designs use multiple trigger signals and distribute the trigger logic into discrete portions of the design~\cite{Zhang2014DeTrust}. However, these new hardware Trojans still assume that the Trojan payload will only be activated under rare conditions. In addition, these Trojans also alter specified design behavior to perform malicious functionality. Tag assisted switching probability analysis \cite{Haider2019Tag} and equivalence checking  based on don't care analysis~\cite{zhang2015veritrust} will identify Trojan design as redundant logic.

More recent research works leverage external or internal don't cares for hardware Trojans. Fern \emph{et al.} used external don't cares introduced by unspecified functions in circuit designs for Trojan design~\cite{fern2015hardware}, which hides malicious functionality behind signals with undeclared behavior. The trigger condition cannot be reached during normal operation. Attackers can force the design into the desired state to launch the attack using illegal test vectors. Since the design space after inserting such a don't care Trojan is a superset of the original design while the designer usually focuses on checking adherence to available specification during verification, detecting such Trojans can be difficult. However, such a Trojan design method does not apply to completely specified circuit designs.  

Hu \emph{et al.} proposed a method that exploits internal don't care conditions for hardware Trojans~\cite{hu2017sdc}, which leveraged internal design state that would never be satisfied during normal operation, e.g., two signals cannot be logical {\tt 1} at the same time as the trigger. Even for the completely specified circuits, there are still numerous internal don't cares for trigger design. An attacker can activate the Trojan through fault injection by over-clocking the design. A successive work \cite{Mahmoud2020Xattack} investigated remote activation of a satisfiability don't care (SDC) Trojan by deploying power-wasting circuits to induce timing faults. This allows an attacker to remotely trigger the SDC Trojan deployed in a logically isolated design partition in a shared FPGA. Another work hides stealthy Trojans into obfuscation logic for IP protection~\cite{hu2019leveraging}. Similarly, the Trojan can be hard to detect due to the fact that there is no open specification for the obfuscation logic.

A recent work employed machine learning (more frequently used as a Trojan detection technique) for automatic Trojan insertion, it created an AI-guided framework for analyzing structural and functional features of existing Trojan designs to train machine learning models for Trojan design generation~\cite{cruz2022MIMIC}. Cruz \emph{et al.}~\cite{cruz2022TRIT-DS} also developed a framework that automatically inserted hardware Trojans in FPGA netlists, using additional LUT to instantiate Trojan templates or leverage dark silicon to realize malicious functionality.

\subsection{Hardware Trojan Detection Techniques}
\label{sec:detecttech}
Existing hardware Trojan detection methods mainly include reverse engineering \cite{Bao2016Reverse}, functional verification \cite{Salmani2012Reducing}, switching probability analysis~\cite{waksman2013fanci}, side-channel analysis \cite{Narasimhan2013MultipleParameter,Huang2018Scalable}, security verification \cite{hu2016detecting,Guo2019Trojan,Jin2012PCH} and machine learning based solutions~\cite{salmani2016cotd,muralidhar2021,yang2021hardware}.

Destructive reverse engineering can be an effective approach for detecting malicious logic insertion when the original design layout is available for comparison~\cite{Bao2016Reverse}. However, it can be a costly process in that reverse engineering recent nano-meter technology chips with billions of transistors requires sophisticated tools and may take years to complete. The destructed chips cannot be deployed any more. In addition, in case only a small number of chips among a large batch contain hardware Trojans, it is impossible to destruct all. However, reverse engineered chip implementation details may provide support for other Trojan detection techniques such as more accurate SCA.

Non-destructive Trojan detection methods are relatively less expensive. These include functional verification, SCA and security verification. The core idea behind functional verification based hardware Trojan detection lies in searching for test vectors that could accelerate the Trojan activation process \cite{Salmani2012Reducing}. Automatic test pattern generation is one such technique to cover low controllability Trojan designs. The MERO \cite{Chakraborty2009MERO} test vector generation method assumes that the Trojan is a malicious circuit activated by low switching probability circuit nodes. It aims to generate a test vector set as small as possible to minimize test effort while maximizing the coverage of Trojan trigger. Another commonly used Trojan detection technique performs switching activity analysis to identify nearly unused circuit in the hardware design~\cite{waksman2013fanci}. This leverages the idea that hardware Trojans are usually only activated under rare events and thus appear as redundancy.

SCA measures the additional path delay, power consumption, EM radiation, and thermal emission to determine the existence of Trojans \cite{Narasimhan2013MultipleParameter}. It usually employs statistical measurements or information theoretic metrics to distinguish small fingerprints. The random noise caused by process variations pose significant challenge on detection precision. There is also research that targets more efficient test vector generation for Trojan detection using SCA \cite{Huang2018Scalable}.

Recently, a large body of research works employ machine learning or deep learning for developing pre-silicon hardware Trojan detection techniques. These techniques leverage various Trojan features, including structural characteristics \cite{muralidhar2021}, functional features \cite{yang2021hardware}, controllability and observability parameters \cite{salmani2016cotd}.
While these advanced Trojan detection techniques can be effective in detecting previously reported Trojan designs, the accuracy depends on the size and completeness of the training data set; they usually require numerous time to tune the parameters in order to reach optimal detection performance. In addition, these techniques may miss a new Trojan, whose feature is not covered by the training data set.

Security verification detects hardware Trojans by formal proof of security properties related to the flow of information, e.g., \emph{confidentiality} and \emph{integrity}. It captures Trojans that cause security property violations \cite{hu2016detecting,Guo2019Trojan}. This type of method can be effective in detecting Trojans that leak secret information or overwrite critical memory locations when triggered. However, such methods can see limitations in detecting Trojans that violate the \emph{availability} property, e.g., causing denial of service. There are also commercial verification tools with taint analysis capabilities such as Siemens EDA \emph{SecureCheck}, Cadence \emph{Jasper}, and Tortuga Logic \emph{Prospect}. These tools can be used to formally verify security properties, e.g., confidentiality and integrity, and detect certain types of hardware Trojans that cause property violations. Such tools are equipped with powerful formal verification engines, which provide better performance and scalability. However, they usually rely on verification engineers to manually specify properties and the quality of the properties specified has a dominant influence on verification outcome. Automatic extracting properties that can cover hard-to-detect design vulnerabilities such as Trojans is an emerging research topic \cite{Deutschbein2022Property}.

\subsection{Hardware Trojan Prevention Techniques}
\label{sec:preventtech}
Embedding Ring Oscillator (RO) infrastructure in the non-critical path of integrated circuits is a frequently used technique for reliability analysis. Inserting additional Trojan circuitry will affect the frequency of the RO and thus provide a fingerprint for Trojan detection. Qu \emph{et al.} \cite{Qu2015RO} developed a hardware Trojan detection method based on RO and proposed an algorithm that used incremental compilation to design RO network. A more recent work designs a 3D RO test infrastructure for detecting Trojans in a 3D die stack \cite{Alhelaly2021RO}. Runtime detection technology provides the last defense line to mitigate Trojans. Hou \emph{et al.} \cite{Hou2018R2D2} proposed an on-chip runtime monitoring scheme for detecting A2-alike Trojans triggered by toggling events.

Logic obfuscation adds additional dummy circuitry out of the specified design state space to prevent attackers from identifying the actual circuit functionality \cite{Dupuis2014Trojan}. Without the correct obfuscation key, it can be difficult to insert a meaningful Trojan. Baumgarten \emph{et al.} \cite{Baumgarten2010Preventing} proposed to prevent malicious manufacturers from inserting Trojans through function reserving. The idea is to insert reconfigurable units, whose functionality is left empty during fabrication and completed (i.e., configured) at the user side.

Split manufacturing takes a different approach to preventing malicious design modification. It splits a design layout into Front End of Line (FEOL) and Back End of Line (BEOL) portions, which will be fabricated by trusted and untrusted foundries respectively. Without information about the BEOL portion, it can be challenging for the untrusted foundry to embed a useful hardware Trojan \cite{Xie2015Split}.

\section{Threat Model}
\label{sec:threat}
We consider typical SoC or IC design scenarios, where designers usually rely on off-the-shelf 3PIP components to lower design effort and cost. We assume that these IP components are designed by vendors of different trust levels and delivered through an untrusted supply chain. Some untrusted IP designers may intentionally add undocumented functionality or make malicious design modifications that are out of the specification of the IP product. These malicious circuitries are inserted into the register transfer level (RTL) or gate-level netlist of IP designs. They are carefully designed and extensively tested so that the Trojan circuitries are triggered only under rare conditions in order to protect them from being identified. We do not account for always-on Trojans, which should expose to routine testing and verification techniques.

We focus on design level Trojans and aim to develop methods for detecting and preventing them during the design phase. We do not consider hardware Trojans that have already been fabricated into a chip, which are out of the scope of CAD and difficult to patch even if being detected. However, our method applies to Trojans residing in FPGA bitstreams thanks to recent advances in bitstream reverse engineering techniques \cite{Ender2020Starbleed} as well as the reconfigurability of FPGA devices.

In this article, we follow a similar threat model adopted by numerous pre-silicon Trojan detection research, e.g., FANCI~\cite{waksman2013fanci} and VeriTrust~\cite{zhang2015veritrust}. Actually, there is no strict line between a hardware Trojan and a hard-to-cover design flaw. A big difference lies in that the former is malicious while the later is unintentional. Our method can be equally effective in identifying hard-to-detect design vulnerabilities activated only under rare conditions.

\section{Preliminaries}
\label{sec:preliminary}
In this section, we cover a few terms and definitions for characterizing hardware Trojans at the level of LUT netlist.

\emph{Definition 1 -- \textbf{Low Switching Signal:} } A \emph{low switching signal} is a signal with significantly biased probability distribution in the observed signal values, i.e., the possibility of the signal taking the value of logical {\tt 0} (or {\tt 1}) approaches $0$ while the possibility of taking the opposite value approaches $1$. In other words, the signal seldom or never changes its logic state. From the perspective of information theory, such signals would create a channel to communicate a significant amount of information when it flips.

\begin{figure}[!htp]
\centering
\includegraphics[width=\columnwidth]{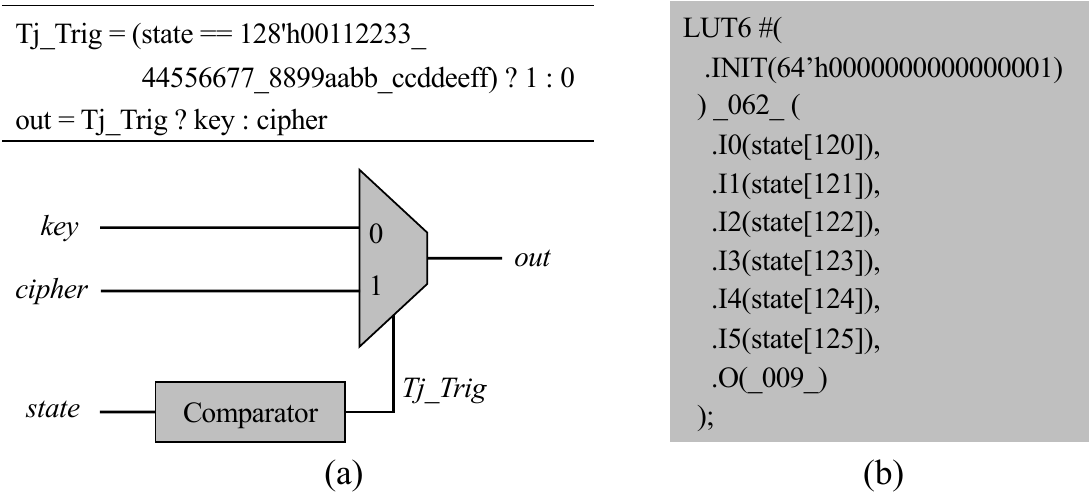}
\caption{The \emph{Trust-Hub} AES-T1000 benchmark, which contains a hardware Trojan that leaks the key to the output when a specific plaintext is observed. (a) Trigger and payload of the Trojan in the benchmark. (b) One of the LUTs for generating the trigger signal in the FPGA netlist of the benchmark.}
\label{fig:t1000}
\end{figure}

Consider the \emph{Trust-Hub}~\cite{Salmani2013Trust-Hub} AES-T1000 benchmark, which contains a hardware Trojan that leaks the \emph{key} when a specific plaintext is observed. From Fig.~\ref{fig:t1000} (a), the \emph{Tj\_Trig} signal only has the possibility of $1/2^{128}$ to take the value of logical {\tt 1} and tends to stay constant during limited testing. However, when \emph{Tj\_Trig} is asserted, it opens up a covert channel to leak the entire key.

Figure~\ref{fig:t1000} (b) shows one of the LUTs for generating the \emph{Tj\_Trig} signal in the FPGA netlist of the benchmark. Even with only six inputs, the output of the LUT does not switch after testing 20,000 random vectors. A number of works start Trojan detection from searching for such low switching activity signals, which are potential triggers \cite{waksman2013fanci}.


\emph{Definition 2 -- \textbf{Low Switching LUT:} } A \emph{low switching LUT} is a LUT, whose output is a low switching signal, i.e., the output of the LUT flips its logic state infrequently like the example in Fig.~\ref{fig:t1000} (b).

\begin{figure}[!htp]
\centering
\includegraphics[width=\columnwidth]{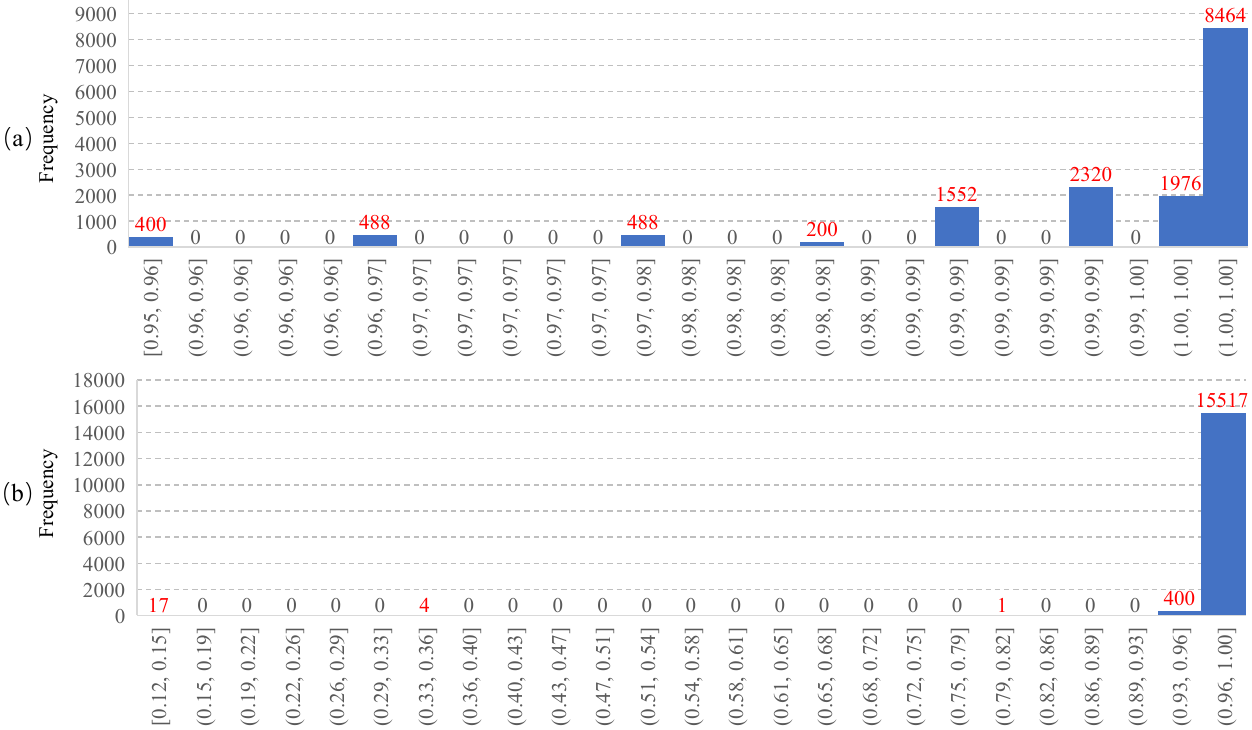}
\caption{Statistics of the entropy values of the initialization vectors in the Trojan-free and Trojan-infected versions of the AES-T1000 benchmark. (a) Frequency counts of the entropy values of the initialization vectors in the Trojan-free benchmark. (b) Frequency counts of the entropy values of the initialization vectors in the Trojan-infected benchmark.} 
\label{fig:entropy}
\end{figure}

The initialization vector of an LUT provides some information for estimating switching probability. For example, an LUT with the initialization vector of {\tt 32'h80000000} has a theoretical switching probability between 1/32 and 1/16 when the address line inputs follow a unified distribution. We employ entropy as a measurement of switching characteristics of LUTs in the AES-T1000 benchmark without and with Trojan, and show the statistics in Fig.~\ref{fig:entropy}. When an initialization vector contains an even number of zeros and ones, the entropy would reach the maximum value of {\tt one}. When the distribution is significantly biased, resulting in much lower switching probability, the entropy should approach {\tt zero}. From Fig.~\ref{fig:entropy} (b), there are 17 extremely low switching LUTs after Trojan insertion, which provides an important clue for Trojan detection. Such features are not available at the levels of RTL code or Boolean gate netlist.


Switching analysis can be effective in detecting hardware Trojans with a single trigger signal. However, more recent Trojans tend to use multiple discrete trigger signals, where each trigger is able to switch normally~\cite{hu2017sdc}. Therefore, we also need to consider the switching characteristics of signal combinations. LUTs offer an ideal granularity to perform such analysis.

\emph{Definition 3 -- \textbf{Low Coverage LUT:} } A \emph{low coverage LUT} is a LUT, where certain address line input combinations are never reached during testing or formal verification. This is usually caused by low switching or don't care conditions in the input lines. 

\begin{figure}[h]
\centering
\includegraphics[width=\columnwidth]{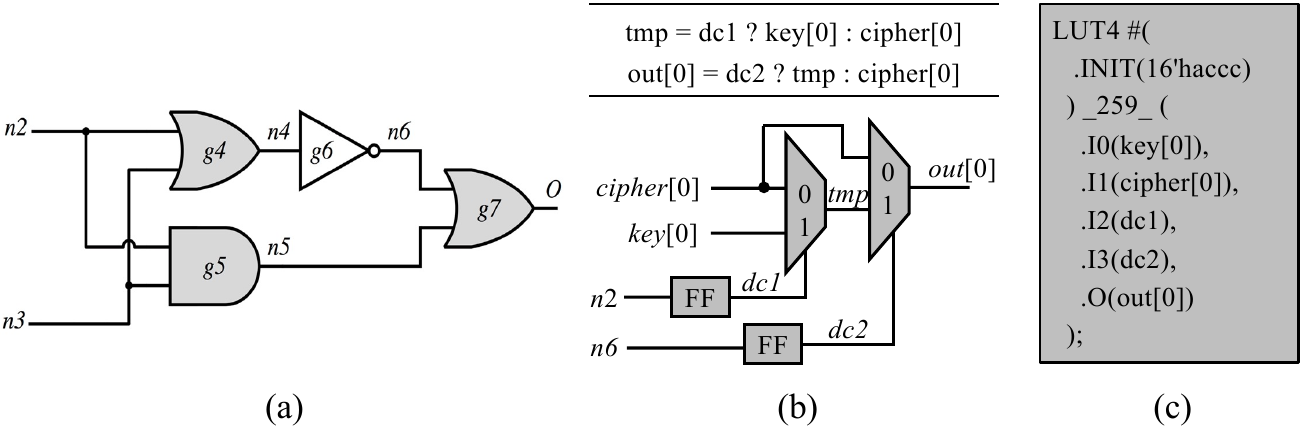}
\caption{A hardware Trojan that is triggered by a satisfiability don't care signal pair. (a) A gate-level netlist that contains an SDC condition. (b) An SDC Trojan implementation. (c) LUT implementation of the Trojan.}
\label{fig:sdc}
\end{figure}

Here, the don't care conditions can be categorized as external and internal. External don't cares are also known as unspecified functionality~\cite{fern2015hardware}, which is usually not accounted for during testing. Satisfiability don't cares are the specific type of internal don't cares that can lead to coverage issues. Such don't cares are caused by signal correlations resulting from reconvergent fanout loops~\cite{hu2017sdc}. Consider the gate-level netlist in Fig.~\ref{fig:sdc}, signals \emph{n2} and \emph{n6} cannot be logical {\tt 1} simultaneously due to path correlation. One can leverage such SDC conditions to design a hardware Trojan as shown in Fig.~\ref{fig:sdc} (b). The Trojan uses two registered SDC signals (i.e., \emph{dc1} and \emph{dc2}) as the triggers, where each trigger is able to switch normally. Thus, existing switching probability analysis based hardware Trojan detection techniques may return a false negative.

When the SDC Trojan design is synthesized into FPGA netlist, the SDC signals \emph{dc1} and \emph{dc2} would connect to a common LUT as shown in Fig.~\ref{fig:sdc} (c). Even if each of them as well as the output of LUT can switch normally, the address line input combinations of the LUT with \emph{dc1} and \emph{dc2} being both logical {\tt 1} can never be reached. This renders a low coverage LUT.

It should be noted that the above concepts are defined from the perspective of testing. The list of \emph{low switching signals}, \emph{low switching LUTs} and \emph{low coverage LUTs} can vary for different simulation runs as we will show in Section~\ref{sec:results}.

\section{Design Flow for Hardware Trojan Detection and Prevention}
\label{sec:flow}
This section starts from a high-level overview of the proposed design flow. It then elaborates the design flow from the aspects of Trojan detection, trigger condition recovery and Trojan prevention respectively.

\subsection{Overview of Our Design Flow}
\begin{figure*}[!htp]
\centering
\includegraphics[width=150mm]{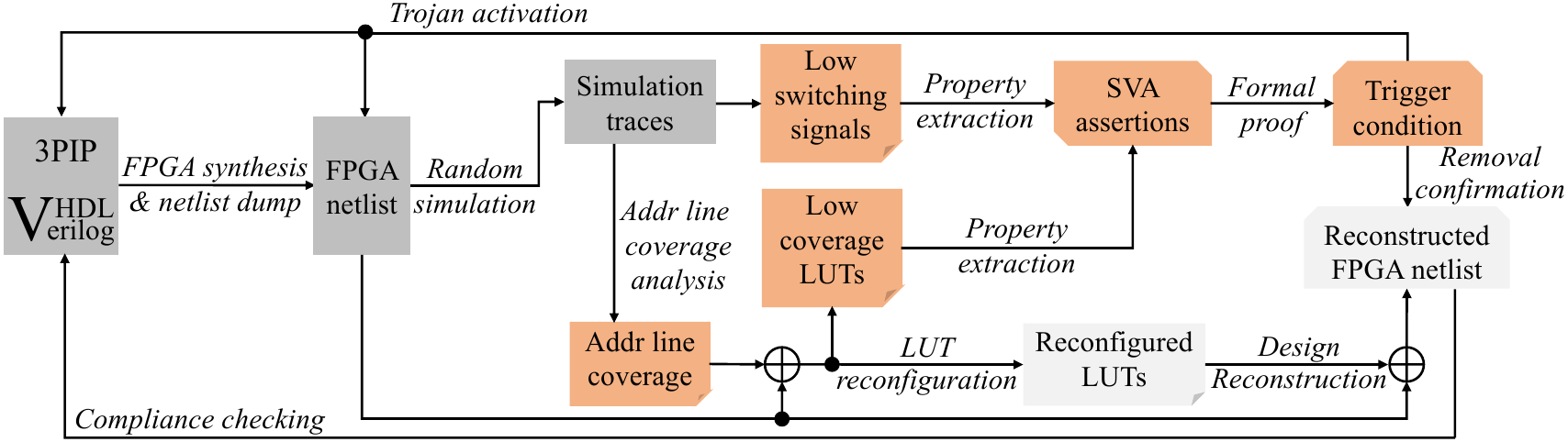}
\caption{The design flow of our method for hardware Trojan detection and prevention.}
\label{fig:flow}
\end{figure*}

Figure~\ref{fig:flow} shows the general flow of the proposed method for detecting and preventing hardware Trojans.

Given a 3PIP in the form of  RTL or gate-level netlist, we first use an FPGA synthesis tool such as \emph{Yosys}~\cite{Shah2019Yosys} and \emph {Xilinx Vivado} to synthesize the IP core. We then dump the synthesized design netlist for further analysis.

We perform random simulation of the synthesized FPGA netlist together with the simulation library for standard cells (e.g., LUTs, MUXes and DFFs) in order to distinguish low switching signals and low coverage LUTs by analyzing the simulation traces. The Trojan trigger is typically a low switching signal while the Trojan payload is usually closely connected to low coverage LUTs. For certain types of Trojans that use multiple discrete triggers, the trigger signals tend to connect to a common LUT after synthesis, rendering a low coverage LUT. We detail the low switching signal and low coverage LUT search process in Section~\ref{sec:detect}.

In our test, we gradually increase the length of simulation trace until the list of low coverage LUTs stabilizes. For the converged set of low coverage LUTs, we assume that the uncovered address line input combinations would never be covered. Afterwards, we automatically extract Trojan related invariant design behaviors from the low switching signals and low coverage LUTs in order to formally check if these switching and coverage related properties always hold. These properties can be specified as SystemVerilog assertions (SVAs) and checked with formal tools. When the proof fails, the formal solver will return a counter example, i.e., the Trojan trigger condition. We describe the property extraction and trigger recovery procedure in Section~\ref{sec:prove}.

We reconfigure the low coverage LUTs that generates the Trojan trigger confirmed during proof step, by flipping the entries corresponding to the uncovered address line input combinations. In case a LUT was used to generate the Trojan trigger before reconfiguration, the Trojan would not be able to activate any more after reconfiguration even if the trigger condition is satisfied. In order to check if the reconfiguration is able to prevent the Trojan from activation, we simulate the FPGA netlists before and after reconfiguration under the trigger condition obtained from formal proof to compare the difference in design behaviors. We illustrate the LUT reconfiguration operation in Section~\ref{sec:prevent}.

\subsection{Hardware Trojan Detection}
\label{sec:detect}
After dumping the synthesized FPGA netlist from \emph{Yosys}~\cite{Shah2019Yosys} or \emph{Vivado}, we write a testbench to perform random simulation on the design netlist. The simulation libraries for different FPGA vendors are provided in \emph{Yosys} distributions. We test the FPGA netlist under a simulation tool such as \emph{ModelSim} and export the simulation trace as an event list\footnote{Using an event list significantly reduces the size of simulation trace.} for successive switching and coverage analysis in order to detect potential Trojan triggers. The core idea is to check the toggling events in the trace to label switched signals and reconstruct the initialization vectors of LUTs. Algorithm~\ref{alg:detection} describes the analysis procedure in detail.
\begin{breakablealgorithm}
\caption{Simulation trace analysis for identifying low switching signals and low coverage LUTs.}
\label{alg:detection}
\begin{algorithmic}[1]
\Require{Design netlist $N$; simulation trace $T$}
\Ensure{The set of low switching signals $S$; the set of low coverage LUTs $L$}

\State{Construct signal list $sList$ from $N$}
\For{$s \in sList$}
\State{$s.value \gets X$}
\State{$s.switch\_tag \gets$ FALSE}
\EndFor

\State{Construct LUT list $cList$ from $N$}
\For{$c \in cList$}
\State{$c.addr \gets X$}
\State{$c.cover \gets 0$}
\EndFor

\State{$time\_step \gets 0$}

\While{$time\_step \leq trace\_length$}
\State{$T \gets$ All toggling events in current time step}

    \For{$t \in T$}
        \For{$s \in sList$}
            \If{$s.signal = t.signal$ and \\
                  ~~~~~~~~~~~~~~~$s.value \neq t.value$}
                \State{$s.switch\_tag \gets$  TRUE}
                \State{$t.updated \gets$ TRUE}
            \EndIf
        \EndFor
    \EndFor
    
    \For{$c \in cList$}
    	\For{$addr\_line \in c.addr$}
	    \If{$addr\_line \in T$ and \\
	        ~~~~~~~~~~~~~~~$addr\_line.updated =$ TRUE}
    	        \State{Update $addr\_line$ in $c.addr$}
	        \State{$c.cover[c.addr] \gets$  TRUE}
	    \EndIf
	\EndFor
    \EndFor
    \State{$time\_step \gets time\_step + 1$}
\EndWhile

\For{$s \in sList$}
    \If{$s.switch\_tag =$ FALSE}
    \State{Add $s$ to low switching signal set $S$}
    \EndIf
\EndFor

\For{$c \in cList$}
    \State{$n \gets$ number of inputs for LUT $c$}
    \If{$c.cover \neq 2^n - 1$}
    \State{Add $c$ to low coverage LUT set $L$}
    \EndIf
\EndFor
\end{algorithmic}
\end{breakablealgorithm}

We first create a signal list from the synthesized design netlist, initialize the value of all signals to $X$ and set their switching tags as FALSE. We also construct a list consisting of all LUTs in the design netlist, assign the initial address lines of all LUTs to $X$ and reset the reconstructed initialization vector $cover$ to zero. For each time step in the simulation trace, we add all toggling events to a set $T$. We check all events in $T$ against the signal list. In case a signal $s$ has a toggling event and its value is updated, we mark the switching tag of $s$ and the update status of $t$ as TRUE. We then check the updates in the address lines of the LUTs. For all address lines of each LUT, if any address line has a toggling event, we update it to calculate a new address. With the updated address, we mark the corresponding bit in the reconstructed initialization vector (i.e., $cover$) as covered.

After traversing the entire simulation trace, we can obtain the updated set of low switching signals $S$ and the set of low coverage LUTs $L$. These signals and LUTs are potential trigger logic of hardware Trojans.

\subsection{Extracting Trojan Related Properties}
\label{sec:prove}
\begin{figure}[!htp]
\centering
\includegraphics[width=\columnwidth]{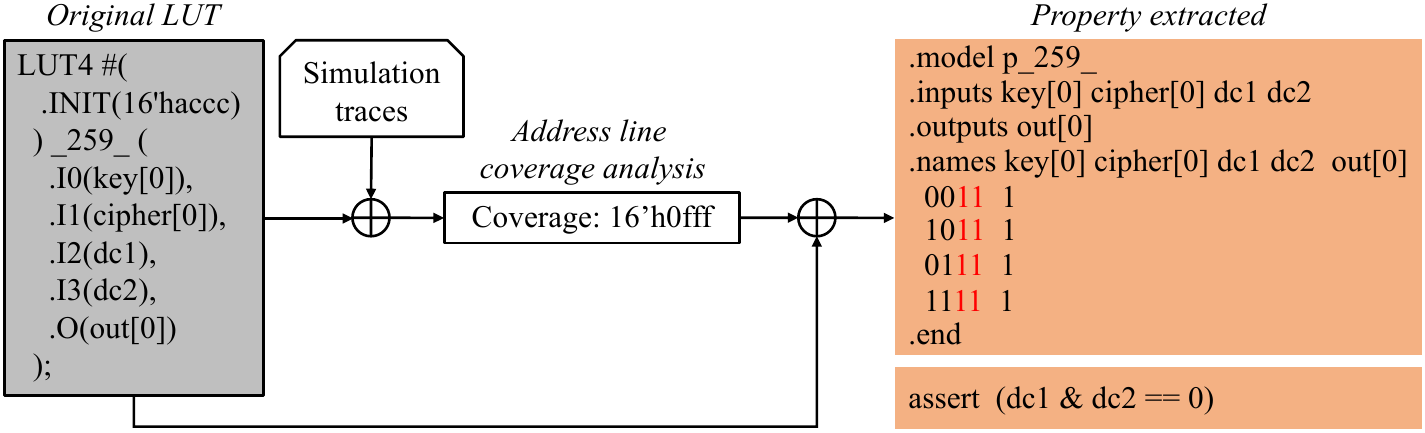}
\caption{Automatic extraction of Trojan related properties.}
\label{fig:property}
\end{figure}

When gradually increasing the simulation time, the set of low switching signals and the set of low coverage LUTs will converge. At this point, we can usually observe some invariant design properties related to switching and coverage. For a low switching signal, we can specify a property, which asserts that the signal stays constant, e.g., \emph{Trojan\_Trigger} should always be logical {\tt 0}. For a low coverage LUT, we can extract a property, which assumes that the uncovered input conditions would never be satisfied. Figure~\ref{fig:property} shows an example of coverage related property extraction.

In the example, the address line input condition statistics stabilize to {\tt 16'h0fff} during coverage analysis. This indicates the highest four address combinations cannot be reached during simulation. These combinations correspond to the cases where the two higher address lines are both logical {\tt 1}. We can create a truth table in BLIF (Berkeley Logic Interchange Format) as shown on the right-hand side of Fig.~\ref{fig:property} for property extraction. The BLIF design can be simplified to the logic equation of \emph{dc}1 \& \emph{dc}2, which yields the hyperproperty that \emph{dc}1 \& \emph{dc}2 should be logical {\tt 0}.

With the extracted property, we can perform assertion based simulation or formal verification to check if the property could be violated. When the property fails to hold during verification, a counter example will be reported. Such a counter example would be the trigger condition, which can cause the low switching or low coverage signals to switch and reveal malicious Trojan behavior.

\subsection{Hardware Trojan Prevention}
\label{sec:prevent}
\begin{figure}[!htp]
\centering
\includegraphics[width=\columnwidth]{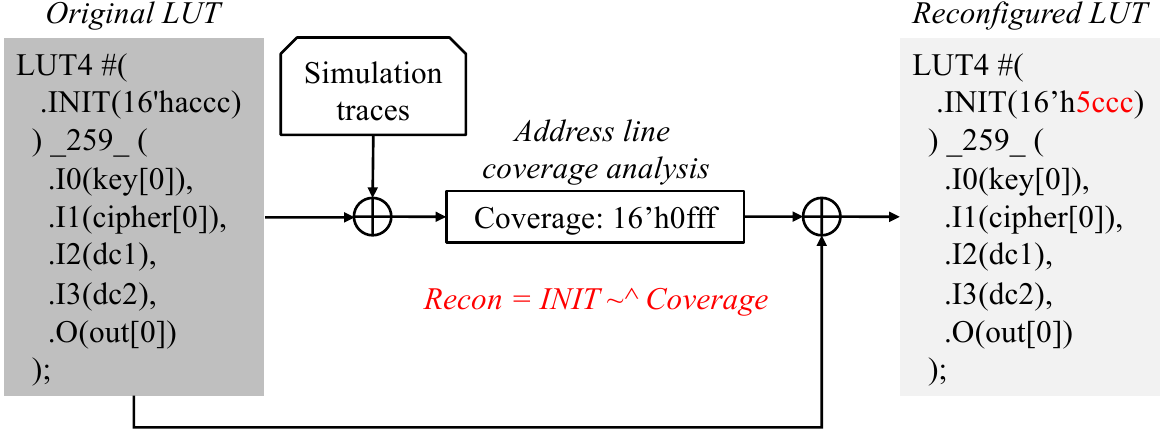}
\caption{Reconfigure LUTs to prevent hardware Trojans from activation.}
\label{fig:reconfig}
\end{figure}

Hardware Trojans usually consist of trigger and payload in order to implement malicious functionality. Thanks to the programmability of FPGA, we can reconfigure the specious LUT identified in the Trojan detection phase to prevent the Trojan from activation. Figure~\ref{fig:reconfig} illustrates this process.

Still consider LUT \_259\_, whose address coverage statistics converges to {\tt 16'h0fff} after extensive random simulation test. From the simulation results, we assume that the two highest address lines cannot be logical {\tt 1} simultaneously. However, the initialization vector of the LUT is {\tt 16'haccc}. It is important to note that the highest four digits have a value of {\tt a} or {\tt 1010} in binary bits, which still has a chance to flip the output of LUT. This indicates that some function is associated with a rare event, i.e., when the two highest address lines are both logical {\tt 1}. Thus, we suspect that some malicious functionality (e.g., the satisfiability don't care Trojan reported in \cite{hu2017sdc}) hides behind the hard-to-reachable condition.

In order to prevent the hidden functionality from activation, we can reconfigure the initialization vector of the LUT as follows, where $\odot$ is the exclusive NOR (XNOR) operator.
\begin{equation*}
\begin{aligned}
Recon &= INIT \odot Coverage \\
           &= \mbox{16'haccc} \odot \mbox{16'h0fff} = \mbox{16'h5ccc}
\end{aligned}
\end{equation*}

After reconfiguration, we can further use a formal equivalence checking tool to check how the updated initialization vector can affect design functionality. In most cases, the formal equivalence checking will fail and return a test vector that allows quick coverage of the low switching probability circuitry. In some cases, the reconfiguration does not change design functionality while mitigating the Trojan, e.g., the external and internal don't care Trojans \cite{fern2015hardware,hu2017sdc,Mahmoud2020Xattack}.

\section{Experimental Results}
\label{sec:results}
This section presents experimental results. We first describe our experimental setup. We report Trojan detection results in Section~\ref{sec:detection}, demonstrate how Trojan trigger condition can be recovered through formal verification in Section~\ref{sec:trigger}, and show how to prevent Trojan from activation by reconfiguring the identified malicious LUTs in Section~\ref{sec:reconfig}.

\subsection{Experimental Setup}
\begin{figure}[!htp]
\centering
\includegraphics[width=\columnwidth]{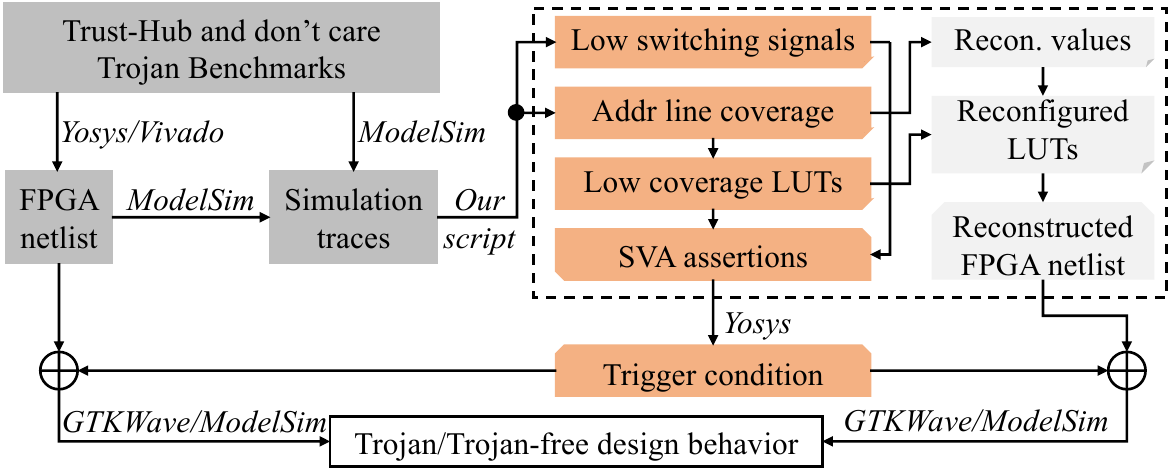}
\caption{Our test flow for hardware Trojan detection and prevention.}
\label{fig:testflow}
\end{figure}

Figure~\ref{fig:testflow} shows our test flow and the design tools used. We employ standard HDL, assertion language and functional verification tools that are familiar to hardware designers as much as possible, especially open source synthesis and verification tools when available. This eliminates the requirement for dedicated automatic test pattern generation or security verification tools and allows security verification to be more closely integrated into functional EDA flow.

As shown in Fig.~\ref{fig:testflow}, we use \emph{Trust-Hub}~\cite{Salmani2013Trust-Hub} and more recently reported don't care Trojan~\cite{hu2017sdc} benchmarks for our tests. We first employ the \emph{Yosys}~\cite{Shah2019Yosys} open source synthesis tool to synthesize the benchmarks and dump the synthesized FPGA netlist for further analysis. We primarily target the Xilinx Virtex-7 and Intel MAX 10 FPGA devices supported by \emph{Yosys}. The corresponding simulation libraries for supported FPGA devices are under the \emph{techlibs} folder in \emph{Yosys} installation directory. When the required simulation library is available, our method also applies to other FPGA devices. For example, we can also export synthesized FPGA netlist from commercial tools such as \emph{Xilinx Vivado}, which supports a wider range of design languages and devices.

We use \emph{ModelSim} to simulate the dumped FPGA netlist in order to collect simulation traces as event lists. We also perform simulation on the original benchmark to check if the FPGA netlist is functionally equivalent to the original design. Such equivalence checking is necessary due to the potential differences in simulated design behaviors between RTL design and FPGA netlist.

We then use our own script to analyze the collected simulation traces in order to identify low switching signals and LUT address line coverage issues. This allows us to pinpoint specious LUTs that have a close connection to Trojan designs. We further use our script to automatically extract invariant properties from the low switching signals as well as low coverage LUTs and specify them as assertion statements. Afterwards, we employ standard functional verification tools such as \emph{Yosys} to formally prove the extracted properties. When a property fails to hold during formal verification, the formal solver will return a counter example that covers the low switching or low coverage condition. For a Trojan related property, the counter example would be the trigger condition. 
Our script also implements low coverage LUT reconfiguration using the method introduced in Section~\ref{sec:prevent} to prevent Trojan from activation. Finally, we use the counter example as a test vector to simulate the original Trojan benchmark as well as the reconfigured FPGA netlist under standard simulation tools such as \emph{GTKWave} to reveal malicious or Trojan-free behavior.

It is important to note that our Trojan detection and prevention method targets LUTs, which represent an ideal granularity for coverage analysis and design reconfiguration. In the following sub-sections, we report test results in Trojan detection, Trojan trigger recovery and Trojan prevention respectively.

\subsection{Trojan Detection Analysis}
\label{sec:detection}
\subsubsection{Test Case 1 -- AES-T1000 Benchmark}
\label{sec:t1000}
\begin{figure}[!htp]
\centering
\includegraphics[width=\columnwidth]{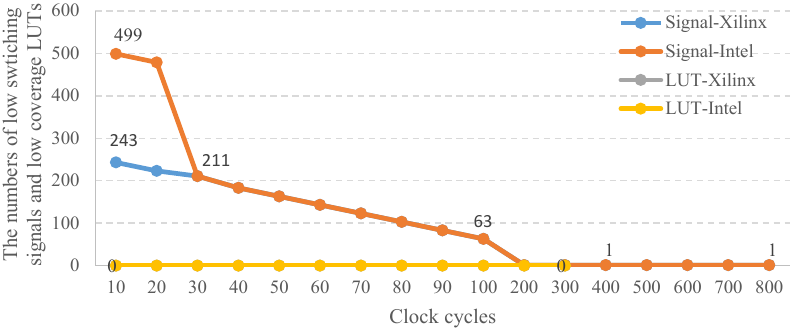}
\caption{Hardware Trojan detection results for the \emph{Top} module in the AES-T1000 benchmark.}
\label{fig:t1000-top}
\end{figure}

We first use the \emph{Trust-Hub} AES-T1000 benchmark to illustrate how our method can detect hardware Trojans by identifying low switching signals and low coverage LUTs. We use the \emph{synth\_xilinx} and \emph{synth\_intel} synthesis scripts in \emph{Yosys}~\cite{Shah2019Yosys} to synthesize the benchmark to FPGA netlists targeting the Xilinx Virtex-7 and Intel MAX10 FPGAs respectively. We then run our Trojan detection method on the \emph{Top} module in synthesized netlist. Figure~\ref{fig:t1000-top} shows the test results.

From Fig.~\ref{fig:t1000-top},  the numbers of low switching signals are 499 and 243 for the Xilinx and Intel FPGA netlists respectively after simulating for 10 clock cycles. The low switching signal counts quickly decrease to only one after testing 400 clock cycles for both netlists. The remaining low switching signal is the Trojan trigger signal, i.e., \emph{Tj\_Trig}. The number of low coverage LUT is zero since there is no LUT in the top module. From the test results, our method precisely captures the Trojan trigger.

\begin{figure}[!htp]
\centering
\includegraphics[width=\columnwidth]{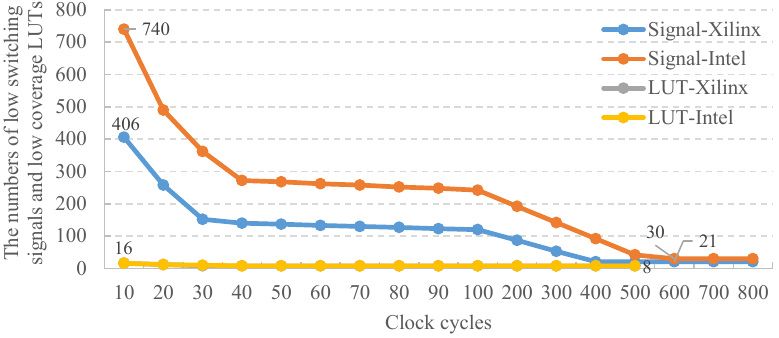}
\caption{Hardware Trojan detection results for the \emph{TSC} module in the AES-T1000 benchmark.}
\label{fig:t1000-tsc}
\end{figure}

We run some additional tests on the \emph{TSC} module. The results are shown in Fig.~\ref{fig:t1000-tsc}.

From Fig.~\ref{fig:t1000-tsc}, the numbers of low switching signals for Xilinx and Intel FPGA netlists converge to 21 and 30 respectively after testing 600 clock cycles. For the Xilinx FPGA netlist, the 21 low switching signals consist of \emph{Tj\_Trig} and a 20-bit counter, which is controlled by \emph{Tj\_Trig}. For the Intel FPGA netlist, the lowest byte of the counter is directly assigned to intermediate wires, adding eight more low switching signals. The remaining specious signals are \emph{Tj\_Trig} and its buffered output. For both FPGA netlists, the numbers of low coverage LUTs stabilize to eight. Each of these eight LUTs has an address line connected to the lowest byte of the \emph{counter}, which does not switch since \emph{Tj\_Trig} is negated. These low switching address lines cause coverage issues in the identified eight LUTs.

\subsubsection{Test Case 2 -- RSA-T200 Benchmark}
\label{sec:t200}
\begin{figure}[!htp]
\centering
\includegraphics[width=\columnwidth]{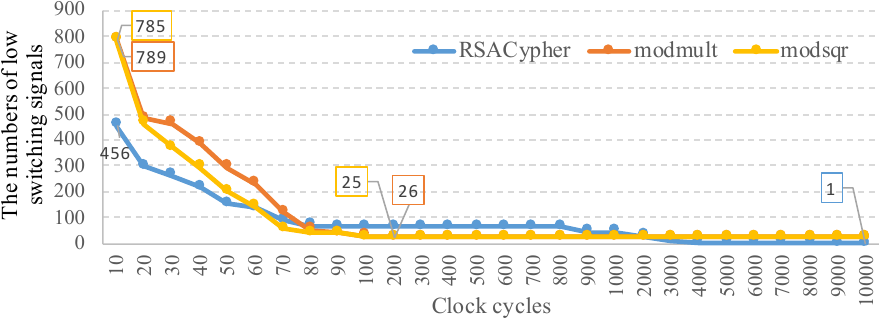}
\caption{Hardware Trojan detection results for the RSA-T200 benchmark.}
\label{fig:t200}
\end{figure}

We then use the \emph{Trust-Hub} RSA-T200 benchmark to demonstrate how our method can work with a commercial FPGA tool. We employ the \emph{Xilinx Vivado} to synthesize the benchmark and run post-synthesis functional simulation, which will produce a simulation model for the benchmark in LUT netlist. We run tests on all the three modules for Trojan detection. Figure~\ref{fig:t200} shows the numbers of low switching signals under different simulation times.

From Fig.~\ref{fig:t200}, the number of low switching signals for the \emph{RSACypher} module converges to {\tt one} after testing 10,000 clock cycles; the numbers of low switching signals for the \emph{modmult} and \emph{modsqr} modules quickly drop from several hundreds to 26 and 25 respectively after simulating for 200 clock cycles. After a closer check of the identified low switching signal sets, the \emph{modmult} and \emph{modsqr} modules each contains 18 unconnected signals that would never switch. Thus, the numbers of specious signals for these two modules reduce to 8 and 7 respectively.  The low switching signal in the \emph{RSACypher} module is exactly the Trojan trigger while the specious signals in the \emph{modmult} and \emph{modsqr} modules are both overflow control logic that would not be reached in normal operation.

\subsubsection{Test Case 3 -- AES SDC Trojan Design}
\label{sec:aes-sdc}
\begin{figure}[!htp]
\centering
\includegraphics[width=\columnwidth]{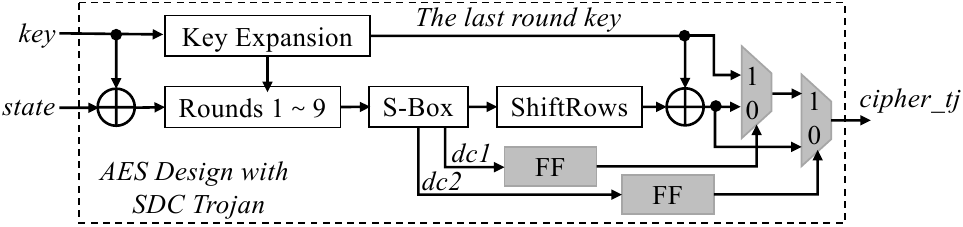}
\caption{AES SDC hardware Trojan design~\cite{hu2017sdc}.}
\label{fig:trojan}
\end{figure}

We further use the AES SDC Trojan proposed in~\cite{hu2017sdc} to demonstrate how our method can detect stealthier don't care Trojans. As shown in Fig.~\ref{fig:trojan}, the Trojan uses an SDC signal pair (i.e., \emph{dc1} and \emph{dc2}), which cannot be logical {\tt 1} simultaneously, from one of the S-Boxes of the final round as two discrete trigger signals. The Trojan multiplexes between the correct encryption result and the last round key. An attacker can force these two SDC signals to be both logical {\tt 1} by simply over-clocking the design to cause a timing failure in one of the flip-flops. This will activate the Trojan and leak the secret round key to the \emph{cipher\_tj} output. The attacker can retrieve the round key by simply plotting the distribution of faulty ciphertext bytes \cite{hu2017sdc,Mahmoud2020Xattack} since the distribution of \emph{cipher\_tj} will see peaks at the round key byte values. This attack method requires much lower effort as compared to typical fault attacks such as differential fault analysis.

The SDC Trojan design is functionally equivalent to its Trojan free baseline since the SDC condition cannot be satisfied during normal operation. Thus, functional verification would not discover the Trojan. In addition, the two discrete SDC trigger signals are both able to flip frequently. Therefore, switching probability analysis would fail to recognize the Trojan triggers as well. The AES SDC Trojan does not show any unique power fingerprint either since the Trojan payload (i.e., the two multiplexers) is consistently switching.

We also use the \emph{synth\_xilinx} and \emph{synth\_intel} scripts to synthesize the Trojan design. We first turn the \emph{flatten} synthesis option off in order to keep the design hierarchy. We perform random simulation of the dumped FPGA netlists for different numbers of timeframes and export the simulation traces as event lists. Afterwards, we use our script to parse the simulation traces to search for \emph{low switching signals} and \emph{low coverage LUTs}.

\begin{figure}[!htp]
\centering
\includegraphics[width=\columnwidth]{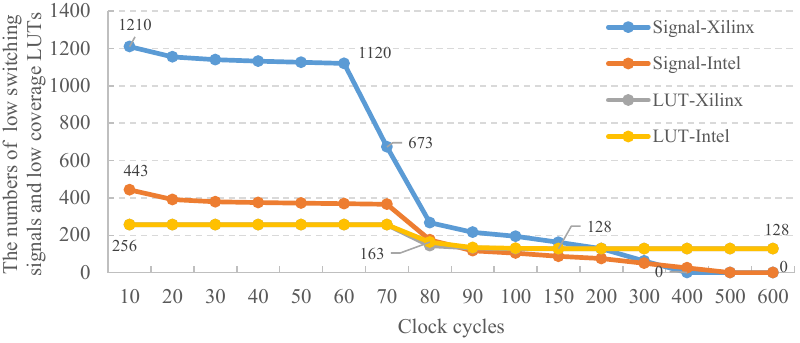}
\caption{AES SDC Trojan design detection results using synthesized Xilinx and Intel FPGA netlists with the \emph{noflatten} option.}
\label{fig:noflatten}
\end{figure}

Figure~\ref{fig:noflatten} shows the detection results for the AES top module. For the Xilinx FPGA netlist, the numbers of low switching signals and specious LUTs are 1210 and 256 respectively after testing 10 clock cycles. The number of specious LUTs converges to 128 only after 100 clock cycles while the number of low switching signals stabilizes to 0 after simulating only 400 clock cycles, taking about 0.25 seconds. This indicates that each signal is able to switch normally and there are 128 specious LUTs. For the Intel FPGA netlist, the numbers of low switching signals and specious LUTs are 443 and 256 respectively after running 10 clock cycles. The number of low switching signals stabilizes to 0 after 500 clock cycles while the number of specious LUTs converges to 128 after simulating 150 clock cycles, taking about 0.20 seconds.

For both design netlists, our Trojan detection method indicates that each signal is able to flip, which matches the design behavior of the AES SDC Trojan. This makes it hard for existing switching probability analysis based Trojan detection methods such as \emph{FANCI}~\cite{waksman2013fanci} to discover the AES SDC Trojan. However, our method identifies 128 specious LUTs. After checking the FPGA netlists, we find that these 128 four-input LUTs multiplex between the 128-bit secret key and the correct encryption results to generate the \emph{cipher\_tj} output; the other two address lines are \emph{dc1} and \emph{dc2}, i.e., the trigger signals. The test results show that our method precisely pinpoints the 128 malicious LUTs associated with the AES SDC Trojan.

It is interesting to note that the Xilinx FPGA netlist leads to a larger number of low switching signals when simulating the same set of test vectors, which is inconsistent with patterns appeared in earlier tests. The \emph{synth\_intel} script by default targets Intel MAX10 devices and uses LUTs with no more than four address lines while the \emph{synth\_xilinx} script sets Virtex-7 FPGAs as the default option and also allows five- and six-input LUTs. Normally, the Intel FPGA netlist would have more intermediate wires and thus a larger number of low switching signals in the beginning. After a close check of the Xilinx FPGA netlist, there are 770 additional wire definitions as compared to the Intel FPGA netlist, which results in a significantly higher number of low switching signals under limited test. However, after simulating a certain number of test vectors, the numbers of low switching signals for both netlists decrease to zero.

\begin{figure}[!htp]
\centering
\includegraphics[width=\columnwidth]{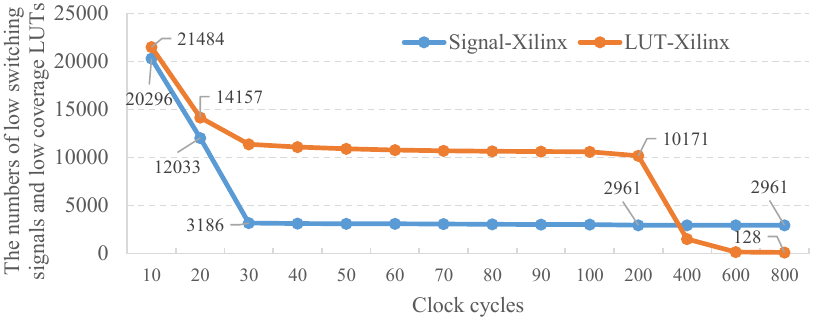}
\caption{Trojan detection results by synthesizing the AES SDC Trojan design into Xilinx FPGA netlist with the \emph{flatten} option.}
\label{fig:flatten}
\end{figure}

We run some additional tests on a \emph{flattened} FPGA netlist created from the AES SDC Trojan using the \emph{synth\_xilinx} script with the \emph{flatten} option. The results are shown in Fig.~\ref{fig:flatten}.

From Fig.~\ref{fig:flatten}, the numbers of low switching signals and low coverage LUTs are 20296 and 21484 respectively after simulating for 10 clock cycles. These are much higher as compared to the corresponding FPGA netlist with design hierarchy. This is because we test the entire \emph{flattened} design rather than the top module only for the \emph{non-flatten} case. The number of low switching signals stabilizes to 2961 after simulating 200 clock cycles. After checking the simulation traces, these low switching signals all stay in high-impedance state, which is caused by design issues instead of hardware Trojan. The number of specious LUTs converges to 128 after testing 800 clock cycles. These 128 specious LUTs implement the same Trojan trigger and payload logic as those in the \emph{non-flatten} version. Similarly, the increase in simulation length is due to the fact that we run detection on the entire design.

\subsubsection{Test Case 4 -- RSA SDC Trojan Design}
\begin{figure}[!htp]
\centering
\includegraphics[width=\columnwidth]{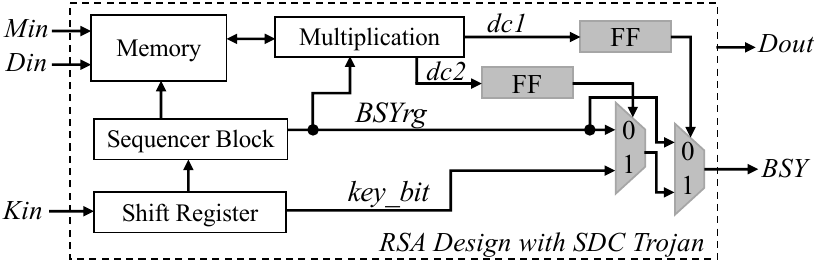}
\caption{RSA SDC hardware Trojan design~\cite{rsa2020}.}
\label{fig:rsa}
\end{figure}

We also use a more recent SDC Trojan design~\cite{rsa2020} inserted into a 1024-bit RSA core as a third test case. Figure~\ref{fig:rsa} illustrates the architecture of the RSA SDC Trojan. Two SDC signals from the multiplication block function as Trojan triggers to multiplex between the most significant bit of the key shift register and the core busy status register \emph{BSYrg} to the \emph{BSY} output. These two SDC signals cannot be logical {\tt 1} at the same time and thus the Trojan will stay inactive during normal operation. We can inject a fault to trigger the Trojan in order to leak the private RSA key bit by bit to the public observable \emph{BSY} output.

\begin{figure}[!htp]
\centering
\includegraphics[width=\columnwidth]{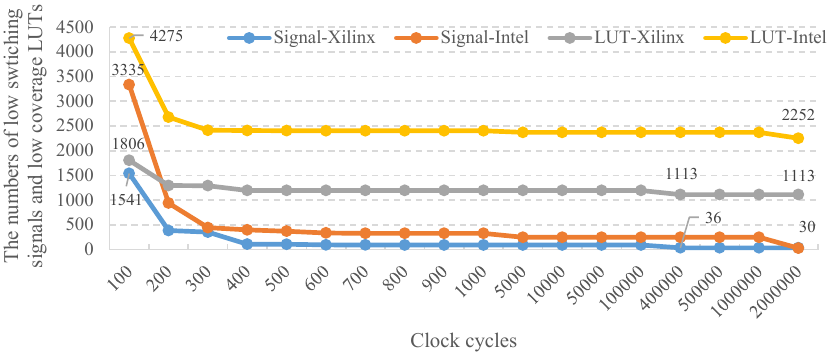}
\caption{RSA SDC Trojan detection results by synthesizing the benchmark into Xilinx and Intel FPGA netlists with the \emph{noflatten} option.}
\label{fig:rsaresults}
\end{figure}

As a first step, we run the \emph{synth\_xilinx} and \emph{synth\_intel} scripts with the \emph{noflatten} option to create Xilinx and Intel FPGA netlists of the RSA SDC Trojan. In our analysis, we choose the top module (i.e., where the Trojan is deployed) in the design hierarchy for illustration. Figure~\ref{fig:rsaresults} shows the Trojan detection results.

For the Xilinx FPGA netlist, the numbers of low switching signals and low coverage LUTs are 1541 and 1806 respectively after simulating 100 clock cycles. These two numbers eventually stabilize to 36 and 1113 respectively after running over 400000 clock cycles. By carefully checking the simulation traces and the FPGA netlist, we find that these 36 low switching signals are either constants or always in high-impedance state. The constants can be excluded from the specious signal list while the high-impedance signals are resulted from design issues. Among the remaining 1113 LUTs, 1024 LUTs are marked as specious due to the common \emph{rst} input, which stays constant after simulation starts; 88 LUTs are connected to high-impedance inputs. The remaining LUT is the SDC Trojan logic that multiplexes the highest bit of the key shift register to the \emph{BSY} output.

By comparison, the numbers of low switching signals and low coverage LUTs for the Intel FPGA netlist are 3335 and 4275 after testing for 100 clock cycles. When testing 2000000 clock cycles, these numbers converge to 30 and 2252 respectively. Similar to the results for Xilinx FPGA netlist, the 30 low switching activity signals are also constants or in high-impedance state. There are 2169 specious LUTs that are driven by constant inputs while another 82 with high-impedance inputs. Again, the remaining single LUT implements the SDC Trojan logic.

\subsubsection{Summary of Trojan Detection Results}
Table~\ref{tab:detection} summarizes the hardware Trojan detection results using several \emph{Trust-Hub} and SDC Trojan benchmarks. 
\begin{table*}
	\caption{Summary of Trojan detection results. The `--'  symbol indicates that no result has been reported for the corresponding method.}
	\begin{center}
	\label{tab:detection}
	\begin{tabular}{| c | c | p{0.48\columnwidth} | p{0.46\columnwidth} | c | c | c | c |}
	\hline
	\# & Benchmark & \centering Trigger condition & \centering Payload & FANCI & VeriTrust & GLIFT &Our Method\\
	\hline
	\multirow{2}{*}{T1} & \multirow{2}{*}{AES-T400} & \multirow{2}{*}{A predefined input plaintext} & Leaks the secret key through a & \multirow{2}{*}{--} & \multirow{2}{*}{--} & \multirow{2}{*}{$\checkmark$} & \multirow{2}{*}{$\checkmark$} \\
	& & & RF signal & & & & \\	
	\hline
	\multirow{3}{*}{T2} & \multirow{3}{*}{AES-T500} &  A predefined sequence of input & A shift register continuously  & \multirow{3}{*}{--} & \multirow{3}{*}{--} & \multirow{3}{*}{--} & \multirow{3}{*}{$\checkmark$} \\
	& & plaintexts & rotates to increases the power & & & & \\
	& & & consumption & & & & \\	
	\hline
	\multirow{2}{*}{T3} & \multirow{2}{*}{AES-T700} & \multirow{2}{*}{A predefined input plaintext} & Leaks the secret key through a & \multirow{2}{*}{--} & \multirow{2}{*}{--} & \multirow{2}{*}{--} & \multirow{2}{*}{$\checkmark$} \\
	& & & covert channel & & & & \\		
	\hline
	\multirow{2}{*}{T4} & \multirow{2}{*}{AES-T800} & A predefined sequence of input & Leaks the secret key through a & \multirow{2}{*}{--} & \multirow{2}{*}{--} & \multirow{2}{*}{--} & \multirow{2}{*}{$\checkmark$} \\
	& & plaintexts & covert channel & & & & \\
	\hline
	\multirow{3}{*}{T5} & \multirow{3}{*}{AES-T900} & \multirow{3}{*}{A specific number of encryptions} & Leaks the secret key trough a  & \multirow{3}{*}{--} & \multirow{3}{*}{--} & \multirow{3}{*}{--} & \multirow{3}{*}{$\checkmark$} \\
	& & & covert CDMA channel in the & & & & \\
	& & & power side-channel & & & & \\	
	\hline
	\multirow{2}{*}{T6} & \multirow{2}{*}{AES-T1000} & \multirow{2}{*}{A specific input plaintext} & Leaks the secret key through a & \multirow{2}{*}{--} & \multirow{2}{*}{--} & \multirow{2}{*}{$\checkmark$} & \multirow{2}{*}{$\checkmark$} \\
	& & & covert channel & & & & \\
	\hline
	\multirow{3}{*}{T7} & \multirow{3}{*}{AES-T1100} & A predefined sequence of input & Leaks the secret key trough a  & \multirow{3}{*}{--} & \multirow{3}{*}{--} & \multirow{3}{*}{$\checkmark$} & \multirow{3}{*}{$\checkmark$} \\
	& & plaintexts & covert CDMA channel in the & & & & \\
	& & & power side-channel & & & & \\
	\hline
	\multirow{3}{*}{T8} & \multirow{3}{*}{AES-T1200} & \multirow{3}{*}{A specific number of encryptions} & Leaks the secret key trough a  & \multirow{3}{*}{--} & \multirow{3}{*}{--} & \multirow{3}{*}{$\checkmark$} & \multirow{3}{*}{$\checkmark$} \\
	& & & covert CDMA channel in the  & & & & \\
	& & & power side-channel & & & & \\
	\hline
	\multirow{3}{*}{T9} & \multirow{3}{*}{AES-T1300} & \multirow{3}{*}{A predefined input plaintext} & Leaks one byte of the AES round  & \multirow{3}{*}{--} & \multirow{3}{*}{--} & \multirow{3}{*}{--} & \multirow{3}{*}{$\checkmark$} \\
	& & & key for each round of the key & & & & \\
	& & &  schedule & & & & \\
	\hline
	\multirow{3}{*}{T10} & \multirow{3}{*}{AES-T1400} & A predefined sequence of input & Leaks one byte of the AES round & \multirow{3}{*}{--} & \multirow{3}{*}{--} & \multirow{3}{*}{--} & \multirow{3}{*}{$\checkmark$} \\
	& & plaintexts &  key for each round of the key & & & & \\
	& & &  schedule & & & & \\
	\hline
	\multirow{2}{*}{T11} & \multirow{2}{*}{AES-T1600} & A predefined sequence of input & Leaks the secret key through a & \multirow{2}{*}{--} & \multirow{2}{*}{--} & \multirow{2}{*}{$\checkmark$} & \multirow{2}{*}{$\checkmark$}\\
	& & plaintexts & RF signal & & & & \\
	\hline
	\multirow{2}{*}{T12} & \multirow{2}{*}{AES-T1700} & \multirow{2}{*}{A specific number of encryptions} & Leaks the secret key information & \multirow{2}{*}{--} & \multirow{2}{*}{--} & \multirow{2}{*}{$\checkmark$} & \multirow{2}{*}{$\checkmark$} \\
	& & & through a RF signal & & & & \\
	\hline
	\multirow{2}{*}{T13} & \multirow{2}{*}{AES-T2000} & A specific sequence of input & Leaks the secret key of AES-128  & \multirow{2}{*}{--} & \multirow{2}{*}{--} & \multirow{2}{*}{--} & \multirow{2}{*}{$\checkmark$} \\
	& & plaintexts & through the leakage current & & & & \\
	\hline
	\multirow{2}{*}{T14} & \multirow{2}{*}{RSA-T100} & \multirow{2}{*}{A predefined input plaintext} & Leaks the secret key trough the & \multirow{2}{*}{--} & \multirow{2}{*}{--} & \multirow{2}{*}{$\checkmark$} & \multirow{2}{*}{$\checkmark$} \\
	& & & cypher output port & & & & \\
	\hline
	\multirow{2}{*}{T15} & \multirow{2}{*}{RSA-T200} & \multirow{2}{*}{A predefined input plaintext} & Replaces the secret key to cause & \multirow{2}{*}{--} & \multirow{2}{*}{--} & \multirow{2}{*}{$\checkmark$} & \multirow{2}{*}{$\checkmark$} \\
	& & & denial of service & & & & \\
	\hline
	\multirow{2}{*}{T16} & \multirow{2}{*}{RSA-T300} & \multirow{2}{*}{A specific number of encryptions} & Leaks the secret key trough the & \multirow{2}{*}{--} & \multirow{2}{*}{--} & \multirow{2}{*}{$\checkmark$} & \multirow{2}{*}{$\checkmark$} \\
	& & & cypher output port & & & & \\
	\hline
	\multirow{2}{*}{T17} & \multirow{2}{*}{RSA-T400} & \multirow{2}{*}{A specific number of encryptions} & Replaces the secret key to cause & \multirow{2}{*}{--} & \multirow{2}{*}{--} & \multirow{2}{*}{$\checkmark$} & \multirow{2}{*}{$\checkmark$} \\
	& & & denial of service & & & & \\
	
	\hline
	\multirow{2}{*}{T18} & \multirow{2}{*}{RS232-T100} & A comparator over 19 signals in & Denial of service by stucking & \multirow{2}{*}{$\checkmark$} & \multirow{2}{*}{$\checkmark$} & \multirow{2}{*}{--} & \multirow{2}{*}{$\checkmark$} \\
	& & the receiver & output signal rec\_readyH at 0 & & & & \\
	\hline
	\multirow{2}{*}{T19} & \multirow{2}{*}{RS232-T300} & A specific number of data & Replaces the 7th bit of all data & \multirow{2}{*}{$\checkmark$} & \multirow{2}{*}{$\checkmark$} & \multirow{2}{*}{--} & \multirow{2}{*}{$\checkmark$} \\
	& & transmission & being transmitted afterward & & & & \\	
	\hline
	\multirow{3}{*}{T20} & \multirow{3}{*}{RS232-T400} & When both transmitted and & Leaks information by replacing 4 & \multirow{3}{*}{$\checkmark$} & \multirow{3}{*}{$\checkmark$} & \multirow{3}{*}{--} & \multirow{3}{*}{$\checkmark$}\\
	& & received data equal to a specific & bits of received data & & & & \\
	& & value &  & & & & \\	
	\hline
	\multirow{2}{*}{T21} & \multirow{2}{*}{RS232-T500} & A specific number of data & Denial of service by stucking & \multirow{2}{*}{$\checkmark$} & \multirow{2}{*}{$\checkmark$} & \multirow{2}{*}{--} & \multirow{2}{*}{$\checkmark$} \\
	& & transmission & output signal xmit\_doneH at 0 & & & & \\
	\hline
	\multirow{2}{*}{T22} & \multirow{2}{*}{RS232-T600} & \multirow{2}{*}{A specific sequence of states} & Leaks information and denial of & \multirow{2}{*}{$\checkmark$} & \multirow{2}{*}{$\checkmark$} & \multirow{2}{*}{--} & \multirow{2}{*}{$\checkmark$} \\
	& & & service & & & & \\	
	\hline
	\multirow{2}{*}{T23} & \multirow{2}{*}{RS232-T700} & \multirow{2}{*}{A specific sequence of states} & Denial of service by stucking  & \multirow{2}{*}{$\checkmark$} & \multirow{2}{*}{$\checkmark$} & \multirow{2}{*}{--} & \multirow{2}{*}{$\checkmark$} \\
	& & & output signal xmit\_doneH at 0 & & & & \\
	\hline
	\multirow{2}{*}{T24} & \multirow{2}{*}{RS232-T900} & \multirow{2}{*}{A specific sequence of states} & Denial of service by blocking & \multirow{2}{*}{$\checkmark$} & \multirow{2}{*}{$\checkmark$} & \multirow{2}{*}{--} & \multirow{2}{*}{$\checkmark$} \\
	& & & data transmission & & & & \\
	
	\hline
	\multirow{3}{*}{T25} & AES & Over-clocking the design to & Leaks the secret key information & \multirow{3}{*}{--} & \multirow{3}{*}{--} & \multirow{3}{*}{--} & \multirow{3}{*}{$\checkmark$}\\
	& SDC & cause a timing failure in order to  & through the Cipher output & & & & \\
	& Trojan & satisfy the SDC condition &  & & & & \\
	\hline
	\multirow{3}{*}{T26} & AES & Deploying power-wasting circuit & Leaks the secret key information & \multirow{3}{*}{--} & \multirow{3}{*}{--} & \multirow{3}{*}{--} & \multirow{3}{*}{$\checkmark$}\\
	& SDC & to cause a timing failure in order & through the Cipher output  & & & & \\
	& Trojan & to satisfy the SDC condition &  & & & & \\
	\hline
	\multirow{3}{*}{T27} & RSA & Over-clocking the design to cause  & Leaks the private RSA key bit by & \multirow{3}{*}{--} & \multirow{3}{*}{--} & \multirow{3}{*}{--} & \multirow{3}{*}{$\checkmark$}\\
	& SDC & a timing failure in order to & bit to the public observable BSY & & & & \\
	& Trojan & satisfy the SDC condition & output & & & & \\
	\hline
	\end{tabular}  
	\label{tab1}
	\end{center}
\end{table*}

From Table~\ref{tab1}, our method can precisely detect the Trojans in \emph{Trust-Hub} benchmarks like other state-of-the-art Trojan detection techniques. In addition, our method can also detect recently reported stealthier don't care Trojans. 

\subsection{Trigger Condition Recovery Analysis}
\label{sec:trigger}
We use the AES-T1000 example to show how our method can automatically extract invariant design properties and recover the Trojan trigger condition through formal proof of the extracted properties. During the Trojan detection phase, there is only one remaining low switching signal (i.e., \emph{Tj\_Trig}) in the top module after simulating 400 clock cycles. Since \emph{Tj\_Trig} stays at logical {\tt 0} in the simulation trace, we specify the following property to assert that it is a constant:
\begin{verbatim}
  assert (Tj_Trig == 0)
\end{verbatim}

We then use \emph{Yosys}~\cite{Shah2019Yosys} to prove the property. The SAT solver in \emph{Yosys} does not automatically search across register boundaries. Thus, it is necessary to trace signal relations across different SAT proofs according to the counter examples returned. Such back tracing can be automated through some lightweight programming to parse the proof results and search the FPGA netlist. We create a flattened version of the benchmark, which allows tracing signal relations across different SAT proofs. Table~\ref{tab:aes-t1000} shows the proof process.
\begin{table}[!ht]
\centering
\caption{The proof process for the AES-T1000 benchmark.}
\label{tab:aes-t1000}
\resizebox{0.48\textwidth}{!}{%
\begin{tabular}{c | c | c | c} \hline
Step & Proof script & Status & Counter example \\
\hline
1 & sat -prove  Tj\_Trig 0 & FAIL & Trigger.Tj\_Trig = 1 \\
\hline
2 & sat -prove \_0779\_ 0 & FAIL & \_0779\_ = 1\\
\hline
3 & sat -prove \_0781\_ 0 & FAIL & state = 00112233445566778899AABBCCDDEEFF \\
\hline
\end{tabular}}
\end{table}

From Table~\ref{tab:aes-t1000}, the first proof step fails and the counter example indicates that \emph{Tj\_Trig} would be logical {\tt 1} when \emph{Trigger.Tj\_Trig} is logical {\tt TRUE}. The \emph{Tj\_Trig} signal is directly connected to \emph{Trigger.Tj\_Trig}, which is generated from \_0779\_ through a non-blocking assignment. Therefore, the proof process then asserts \_0779\_ is constantly logical {\tt 0} during the second step. The \_0779\_ signal is evaluated in a conditional branch statement block, where \_0779\_ would be logical {\tt 1} only if \_0781\_ is logical {\tt TRUE}. Thus, a property related to \_0781\_  is proved in the third step. Our method traces back across two register boundaries and successfully recovers the Trojan trigger condition, i.e., the last counter example.

For the AES SDC Trojan test case, there are 128 remaining low coverage LUTs, whose recovered address line coverage is {\tt 16'h0fff}. This indicates that the highest four entries of these LUTs cannot be reached. We can extract the following property for the two higher address lines \emph{dc1} and \emph{dc2}:
\begin{verbatim}
  assert (dc1 & dc2 == 0)
\end{verbatim}

Similarly, we create a flattened version of the AES SDC Trojan benchmark to allow easier tracing of signal relations. Table~\ref{tab:aes-sdc} shows the proof process. The major difference from the AES-T1000 example is that the property involves two signals. Thus, we need to prove a property instance for each signal in every step.
\begin{table*}[!ht]
\centering
\caption{The proof process for the AES SDC Trojan benchmark.}
\label{tab:aes-sdc}
\begin{tabular}{c | c | c | c} \hline
Step & Proof script & Status & Counter example \\
\hline
\multirow{2}{*}{1} & sat -prove  dc1 0 -set dc2 1 & FAIL & AES.rf.S4\_1.S\_0.dc1 = 1,  AES.rf.S4\_1.S\_0.dc2 = 1 \\
\cline{2-4}
 & sat -prove  dc2 0 -set dc1 1 & FAIL & AES.rf.S4\_1.S\_0.dc1 = 1,  AES.rf.S4\_1.S\_0.dc2 = 1 \\
\hline
\multirow{2}{*}{2} & sat -prove  \_0764\_ 0 -set \_0765\_ 1 & FAIL & \_0764\_ = 1,  \_0765\_ = 1 \\
\cline{2-4}
 & sat -prove  \_0765\_ 0 -set \_0764\_ 1 & FAIL & \_0764\_ = 1,  \_0765\_ = 1 \\
\hline
\multirow{2}{*}{3} & sat -prove AES.rf.S4\_1.S\_0.in[7] 0 -set AES.rf.S4\_1.S\_0.n885 1 & SUCCESS & N.A. \\
\cline{2-4}
 & sat -prove AES.rf.S4\_1.S\_0.n885 0 -set AES.rf.S4\_1.S\_0.in[7] 1 & SUCCESS & N.A. \\
\hline
\end{tabular}
\end{table*}

From Table~\ref{tab:aes-sdc}, the property is formally proved after tracing back across two register boundaries. In other words, \emph{dc1} and \emph{dc2} cannot be logical {\tt 1} simultaneously, which agrees with the fact that they compose an SDC condition.

\subsection{LUT and Design Reconstruction Analysis}
\label{sec:reconfig}
We use the AES-T1000 benchmark to demonstrate how our method can prevent Trojan from activation leveraging the reconfigurability of FPGA. 

Our reconfiguration policy is to prevent the identified trigger signal from flipping even when the trigger condition is satisfied. Before reconfiguration, a formal proof is performed to confirm the output signal generated by the target LUT is the Trojan trigger and recover the trigger condition. Those covered bits in the LUT initialization vector during simulation should not be the activation condition of the Trojan and thus can maintain their values.  Those uncovered bits would be the trigger condition and thus their values should be flipped. That is the reason why we perform an exclusive NOR of the initialization value and the address coverage. Since the trigger signal should stay in a non-active state during normal operation, localized lightweight reconfiguration to prevent the trigger signal from flipping under trigger condition would not change normal functionality. Figure 16 has shown that reconfiguration does not affect the cipher output.

\begin{figure*}[!htp]
\centering
\includegraphics[width=160mm]{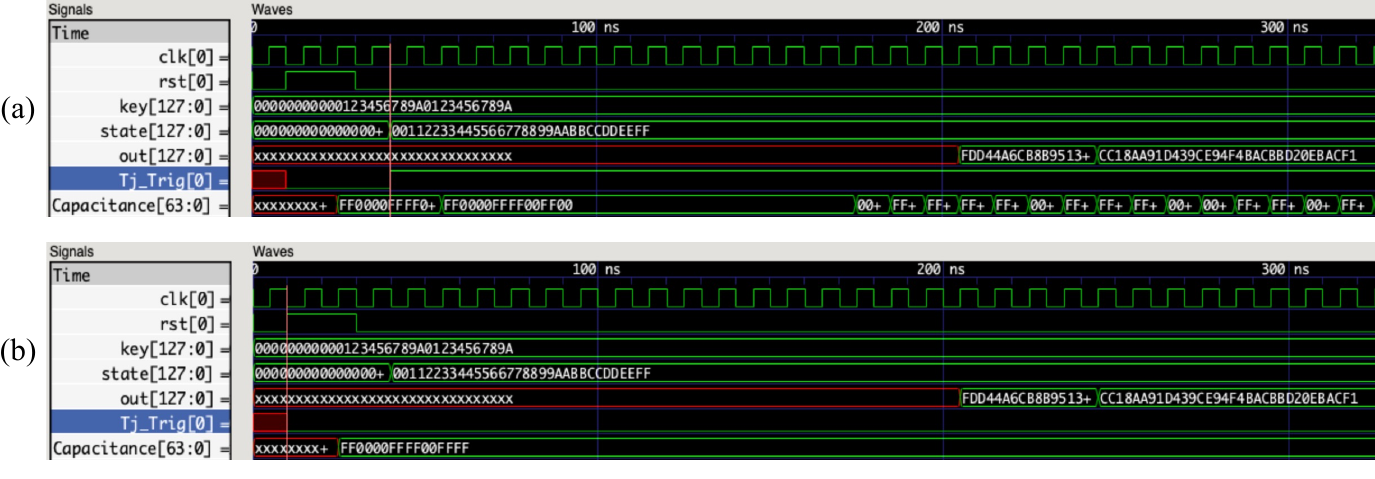}
\caption{Simulation results of the AES-T1000 FPGA netlist before and after reconfiguration with the counter example from formal proof. (a) Simulation result before reconfiguration. (b) Simulation result after reconfiguration.}
\label{fig:wave}
\end{figure*}

Now that \emph{Tj\_Trig} has been identified as the Trojan tigger, we can locate and reconfigure the LUT that generates this malicious signal in the FPGA netlist. We then simulate the FPGA netlists before and after reconfiguration with the counter example (i.e., the Trojan trigger condition) from formal proof. Figure~\ref{fig:wave} shows the simulation results using \emph{GTKWave}.

From Fig.~\ref{fig:wave}, the \emph{Tj\_Trig} signal in the original FPGA netlist will be asserted when the trigger condition is satisfied. Consequently, the secret key will leak to the \emph{Capacitance} output. By comparison, the \emph{Tj\_Trig} signal in the reconfigured netlist will stay low even when the trigger event is observed and there is no switching activity in the \emph{Capacitance} output. In addition, the local lightweight reconfiguration does not have an effect on the ciphertext output.

Although results have shown the effectiveness of our method in detecting and mitigating hardware Trojans, it still has some limitations. The proposed method targets the design phase and requires access to the design source or netlist; it cannot be used for detecting Trojans in a fabricated chip, which is out of the scope of this work. Some Trojans with floating outputs will be optimized away during logic synthesis; our method will indicate the synthesized design netlist is Trojan-free. In addition, there is possibility that coverage analysis may miss Trojans activated by multiple discrete trigger signals that are connected to completely different LUTs.

\section{Conclusion}
\label{sec:conclusion}
In this article, we have demonstrated a design flow for detecting and preventing hardware Trojans in 3PIPs. We leverage standard FPGA synthesis, simulation and verification tools to dump synthesized design netlist, run random simulation and check invariant design properties. We perform switching and coverage analysis at the level of LUTs in order to pinpoint stealthy hardware Trojans hidden behind low switching probability trigger signals or hard-to-reach don't care conditions. The proposed design flow automatically extracts Trojan related properties from invariant design behaviors observed during random simulation and formally checks these properties in order to recover Trojan trigger conditions. We reconfigure specious LUTs with low switching or coverage behaviors to mitigate malicious design functionality. As compared to other design level Trojan detection methods that target Boolean gates or RTL design, LUT is an ideal target for switching and coverage analysis, property extraction as well as design reconfiguration. Our method also sees significant advantage in identifying the more recent don't care Trojans.

\ifCLASSOPTIONcaptionsoff
  \newpage
\fi

\bibliographystyle{IEEEtran}

\bibliography{IEEEabrv,./reference}

\begin{IEEEbiography}[{\includegraphics[width=1in,height=1.25in,clip,keepaspectratio]{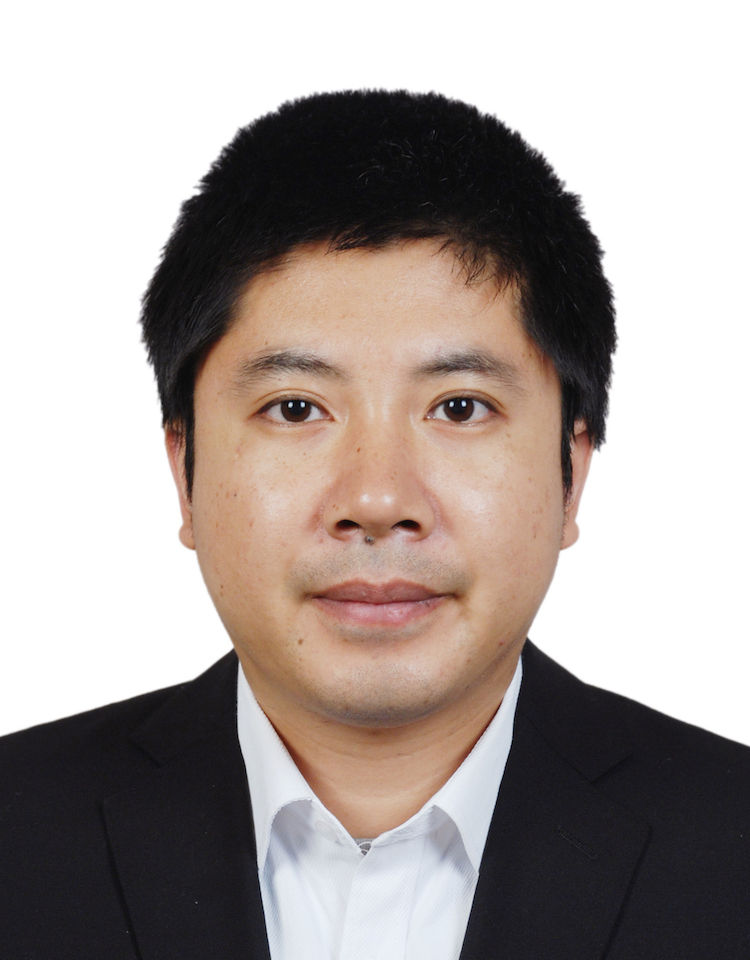}}]{Wei Hu (M'17)} is currently an associate professor with the School of Cybersecurity, Northwestern Polytechnical University. He got his PhD degree in Control Science and Engineering from the same university. His research interests are in hardware security, cryptography, formal security verification, logic and high-level synthesis, formal methods and reconfigurable computing.
 
Dr. Hu serves as the Guest Associate Editor of IEEE Transactions on Computer-Aided Design of Integrated Circuits and Systems. He has been an Organizing Committee member of IEEE International Symposium on Hardware Oriented Security and Trust and Asian Hardware Oriented Security and Trust Symposium since 2017. He was the Technical Program Co-Chair of 2019 Asian Hardware Oriented Security and Trust Symposium and Technical Program Committee Member of ICCD, ASAP and CFTC. He published over 70 papers in peer-reviewed journals and conferences, 2 books and 3 patents.
\end{IEEEbiography}

\begin{IEEEbiography}[{\includegraphics[width=1in,height=1.25in,clip,keepaspectratio]{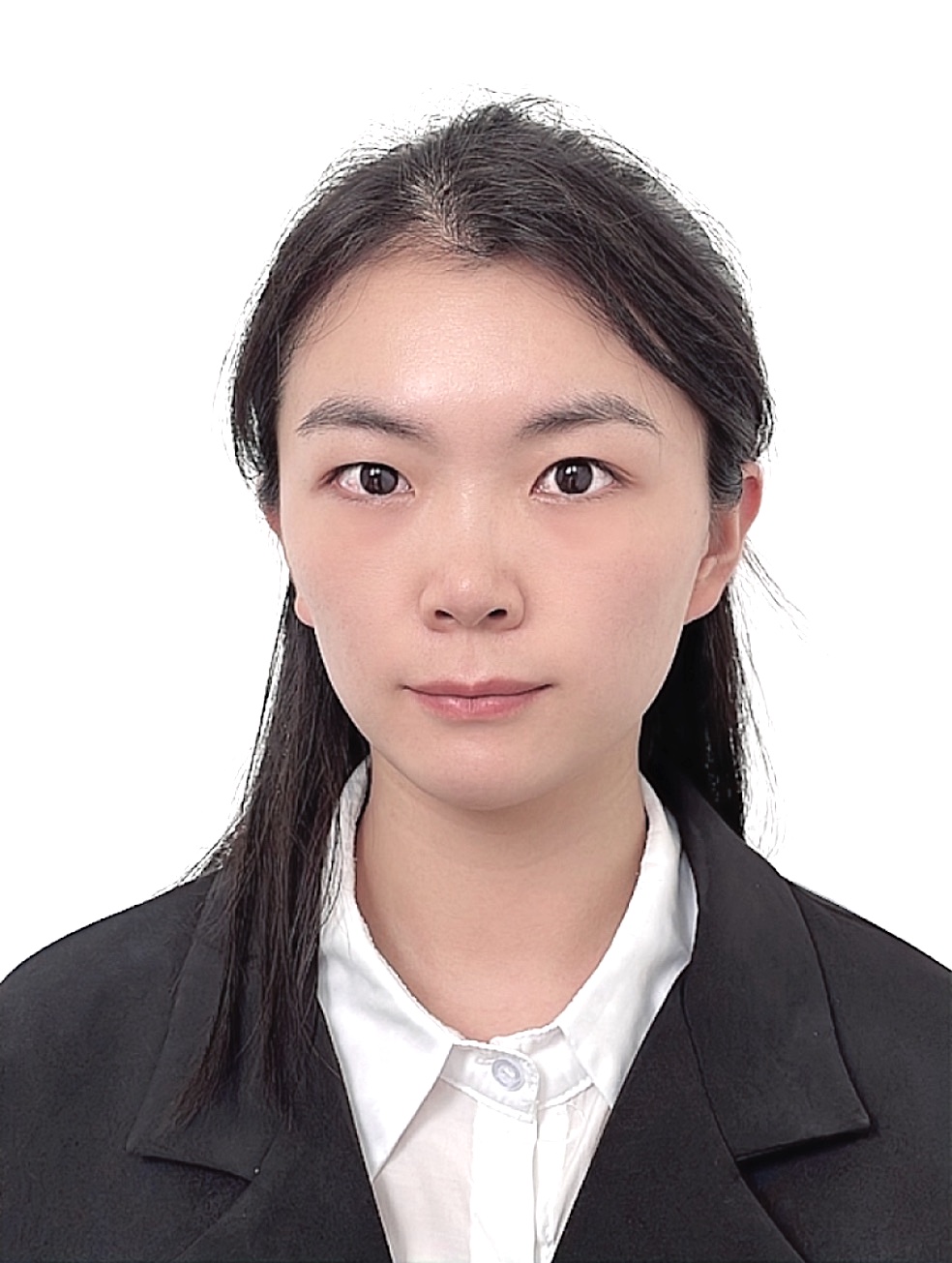}}]{Beibei Li} received the B.S. degree in software engineering from Harbin University in 2017, the M.S. degree in computer technology from Northwest Normal University in 2020. She is currently pursuing the Ph.D. with the School of Cybersecurity, Northwestern Polytechnical University. 

Her research interests are in various aspects of hardware security, including secure architecture, formal security verification, side channel analysis and hardware Trojan detection.
\end{IEEEbiography}

\begin{IEEEbiography}[{\includegraphics[width=1in,height=1.25in,clip,keepaspectratio]{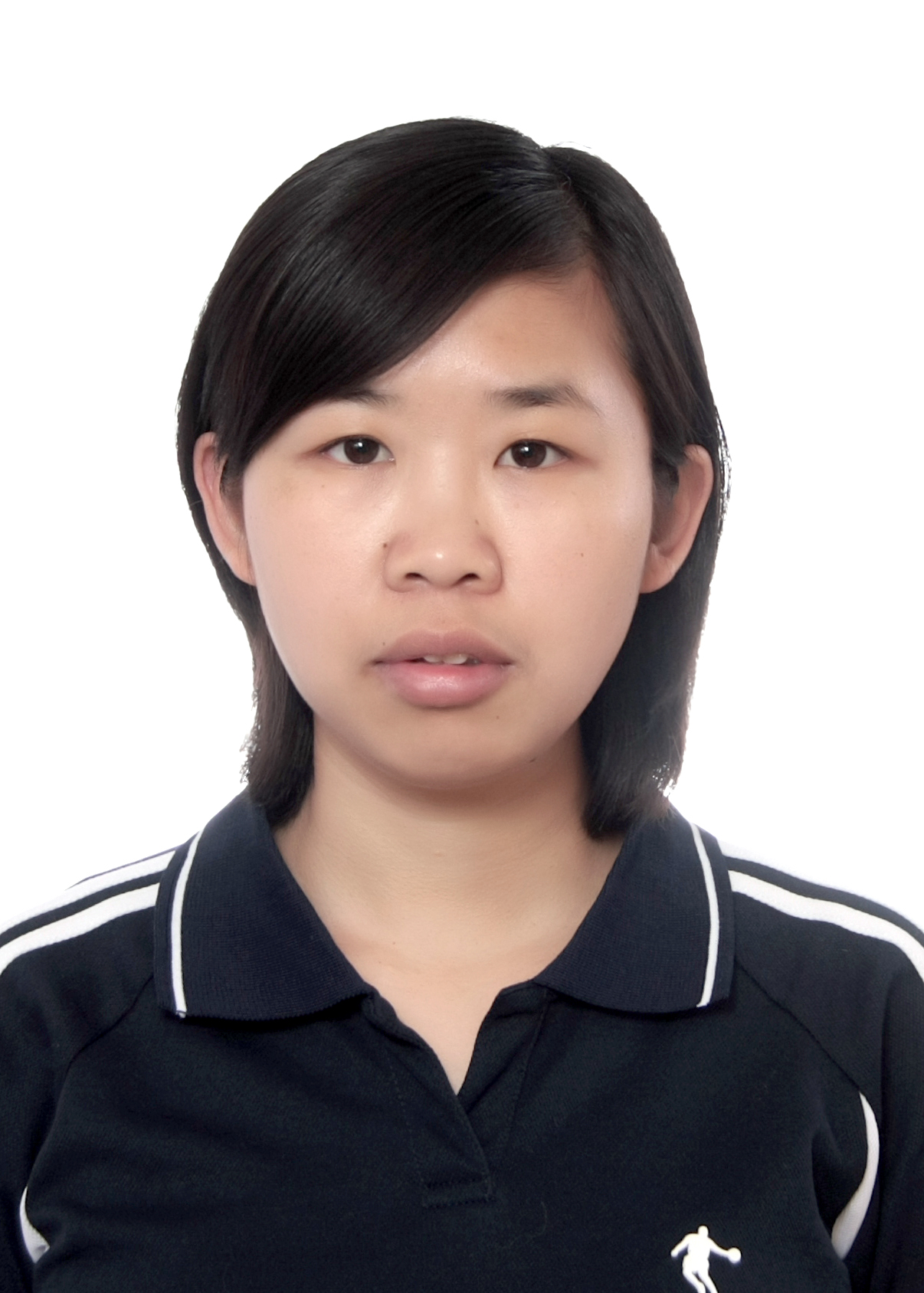}}]{Lingjuan Wu} got her PhD degree in Microelectronics and Solid State Electronics from the Peking University in 2013. She visited the University of California, San Diego as a research scholar from 2010 to 2012.

She is currently an associate professor with the College of Informatics, Huazhong Agricultural University. Her research interests are in hardware security, including secure architecture, formal security verification, side channel analysis and hardware Trojan detection.
\end{IEEEbiography}

\begin{IEEEbiography}[{\includegraphics[width=1in,height=1.25in,clip,keepaspectratio]{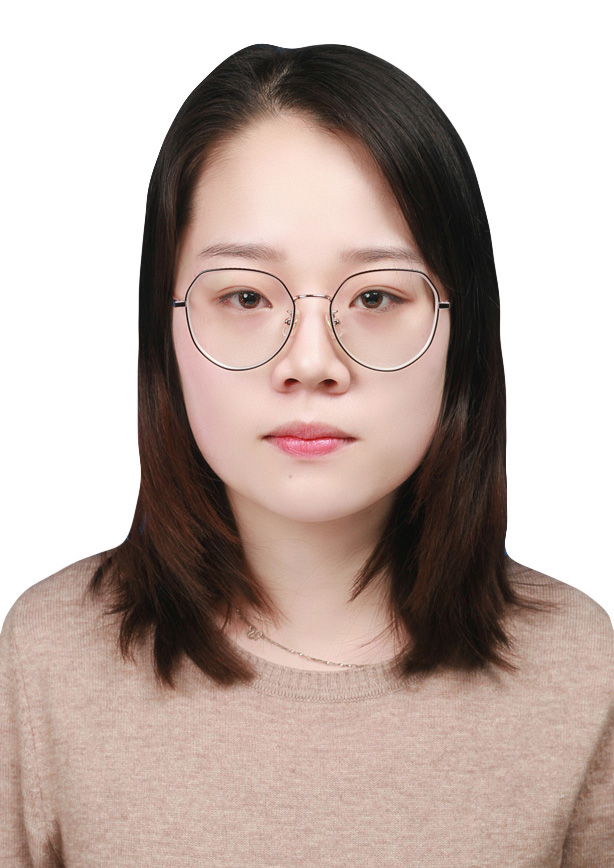}}]{Yiwei Li} is currently an undergraduate student with the School of Cybersecurity, Northwestern Polytechnical University, China. 

Her current research interests include hardware security verification, hardware Trojan detection, formal methods and reconfigurable computing.
\end{IEEEbiography}

\begin{IEEEbiography}[{\includegraphics[width=1in,height=1.25in,clip,keepaspectratio]{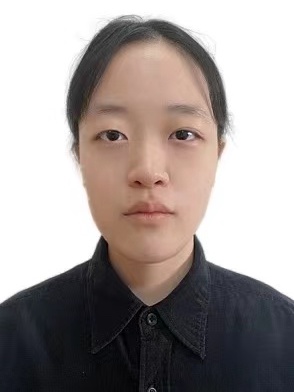}}]{Xuefei Li} Xuefei Li received her B. E. degree from the Changsha University of Science \& Technology, China, in 2019. She is currently a master's student with the School of Cybersecurity, Northwestern Polytechnical University, China. 

Her current research interests lie in hardware security, including hardware security verification, hardware Trojan detection, side channel analysis and cryptography.
\end{IEEEbiography}

\begin{IEEEbiography}[{\includegraphics[width=1in,height=1.25in,clip,keepaspectratio]{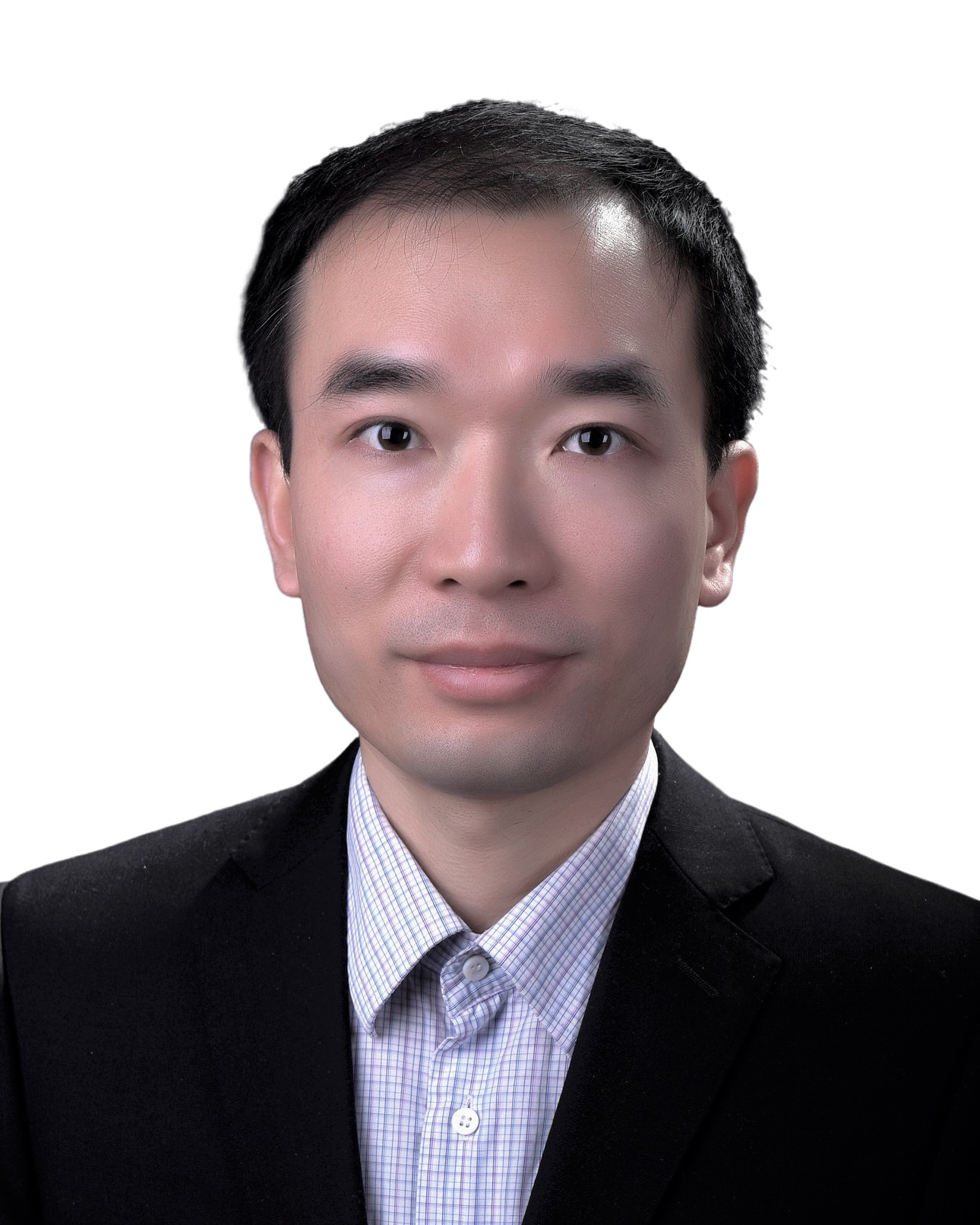}}]{Liang Hong} (M'19) received his B.S. degree in Computer Science and Technology and M.S. degree in Electrical Engineering both from the Harbin Institute of Technology, China in 2001 and 2003, respectively, and the PhD degree in Computer Science from the Huazhong University of Science \& Technology, China in 2006. 

He is currently an associate professor with the School of Cybersecurity, Northwestern Polytechnical University, China. His current research interests include hardware security, cyber security and wireless communications.
\end{IEEEbiography}

\end{document}